\documentclass{aa}  
\usepackage{graphicx}
\usepackage{txfonts}
\usepackage{hyperref}
\usepackage{color}

\def\dsct{$\delta$~Scuti }

\def\Msun{$M_{\odot}$}

\def\ang{$\mathring{A}$}
\def\Teff{\ensuremath{T_{\mathrm{eff}}}}
\def\cd{d$^{\rm -1}$}
\def\logg{\ensuremath{\log g}}
\def\vmic{$\upsilon_{\mathrm{mic}}$}

\def\vsini{\ensuremath{{\upsilon}_{\rm e}\sin i}}
\def\kms{$\mathrm{km\,s}^{-1}$}

\begin{document}

   \title{$\beta$\,Cas: the first $\delta$\,Scuti star with a dynamo magnetic field\thanks{Based on data collected by the BRITE Constellation satellite mission, designed, built, launched, operated and supported by the Austrian Research Promotion Agency (FFG), the University of Vienna, the Technical University of Graz, the University of Innsbruck, the Canadian Space Agency (CSA), the University of Toronto Institute for Aerospace Studies (UTIAS), the Foundation for Polish Science \& Technology (FNiTP MNiSW), and National Science Centre (NCN). Also based on observations obtained at the Telescope Bernard Lyot (USR5026)  operated by the Observatoire Midi-Pyrénées, Universit\'e de Toulouse (Paul Sabatier), Centre National de la Recherche Scientifique (CNRS) of France}}

   \author{K. Zwintz\inst{1} \and
          C. Neiner\inst{2} \and
          O. Kochukhov\inst{3} \and
          T. Ryabchikova\inst{4} \and
          A. Pigulski\inst{5} \and
          M. M\"ullner\inst{1} \and
          T. Steindl\inst{1} \and
          R. Kuschnig\inst{6} \and
          G. Handler\inst{7} \and
          A. F. J. Moffat\inst{8} \and 
          H. Pablo\inst{9} \and 
          A. Popowicz\inst{10} \and
          G. A. Wade\inst{11}
          }

   \institute{Institute for Astro- and Particle Physics, Universit\"at Innsbruck, Technikerstrasse 25, A-6020 Innsbruck\\
              \email{konstanze.zwintz@uibk.ac.at} 
\and LESIA, Paris Observatory, PSL University, CNRS, Sorbonne
Universit\'e, Universit\'e de Paris, 5 place Jules Janssen, 92195 Meudon, France
\and Department of Physics and Astronomy, Uppsala University, Box 516, 75120 Uppsala, Sweden
\and Institute of Astronomy, Russian Academy of Sciences (RAS), Pyatnitskaya 48, 119017 Moscow, Russia
% #5 - Wroclaw, Andrzej
\and 
Instytut Astronomiczny, Uniwersytet Wroclawski, ul. Kopernika 11, PL-51-622 Wroclaw, Poland
% #6 TU Graz - Rainer
\and 
Institut f\"ur Kommunikationsnetze und Satellitenkommunikation, Technical University Graz, Inffeldgasse 12, A-8010 Graz, Austria
\and 
% #7 Gerald
Nicolaus Copernicus Astronomical Center, ul. Bartycka 18, 00-716 Warsaw, Poland     
\and 
% #8 Tony Moffat
D\'epartement de physique, Universit\'e de Montr\'eal, CP 6128, Succursale Centre-Ville, Montr\'eal, Qu\'ebec H3C 3J7, Canada; Centre de Recherche en Astrophysique du Qu\'ebec (CRAQ), Montr\'eal, Qu\'ebec H3C 3J7, Canada
\and
% #9 Bert Pablo
American Association of Variable Star Observers, 49 Bay State Road, Cambridge, MA 02138, USA
% #10 Adam Popowicz
\and 
Silesian University of Technology, Department of Electronics, Electrical Engineering and Microelectronics, Akademicka 15, 44-100 Gliwice, Poland
\and 
% #11 - Kingston, Gregg
Department of Physics and Space Science, Royal Military College of Canada, PO Box 17000, Stn Forces, Kingston, K7K 7B4 Ontario, Canada
}

   \date{Received; accepted}

% \abstract{}{}{}{}{} 
% 5 {} token are mandatory
 
  \abstract
   {F type stars are characterised by several physical processes such as different pulsation mechanisms, rotation, convection, diffusion, and magnetic fields. The rapidly rotating $\delta$ Scuti star $\beta$\,Cas can be considered as a benchmark star to study the interaction of several of these effects.}
   {We investigate the pulsational and magnetic field properties of $\beta$\,Cas. We also determine the star's apparent fundamental parameters and chemical abundances.}
   {Based on photometric time series obtained from three different space missions (BRITE-Constellation, SMEI, and TESS), we conduct a frequency analysis and investigate the stability of the pulsation amplitudes over four years of observations. We investigate the presence of a magnetic field and its properties using spectropolarimetric observations taken with the Narval instrument by applying the Least Square Deconvolution and Zeeman Doppler Imaging techniques. }
   {$\beta$\,Cas shows only three independent p-mode frequencies down to the few ppm-level; its highest amplitude frequency is suggested to be a $n=3$, $\ell = 2$, $m=0$ mode. Its magnetic field structure is quite complex and almost certainly of a dynamo origin. $\beta$\,Cas' atmosphere is slightly deficient in iron peak elements and slightly overabundant in C, O, and heavier elements.} 
   {Atypically for $\delta$ Scuti stars, we can only detect three pulsation modes down to exceptionally low noise levels for $\beta$\,Cas. The star is also one of very few $\delta$ Scuti pulsators known to date to show a measurable magnetic field, and the first $\delta$ Scuti star with a dynamo magnetic field. These characteristics make $\beta$\,Cas an interesting target for future studies of dynamo processes in the thin convective envelopes of F-type stars, of the transition region between fossil and dynamo fields, and the interaction between pulsations and magnetic field. }


   \maketitle
   \titlerunning{$\beta$\,Cas: pulsations and magnetic field}
   \authorrunning{K. Zwintz et al.}
%
%________________________________________________________________

\section{Introduction}

%\subsection{Context}

Stellar evolution is influenced by the interaction of several physical processes such as pulsations, rotation, magnetic fields, convection, and diffusion. The lifetimes of stars changes depending on how strong is the impact of these effects. 
The exact description and theoretical representation of all interacting physical processes remains one of the great challenges in stellar astrophysics. Through the analysis of suitable benchmark objects, we aim to learn more about these physical interactions and improve our theoretical understanding.

F-type stars can show four types of pulsations: (i) $\delta$ Scuti-type p-modes driven by the $\kappa$-mechanism acting in the \ion{He}{ii} ionisation zone \citep{pamyatnykh2000}, (ii) $\gamma$ Doradus type g-modes that are caused by the convective flux blocking mechanism \citep{guzik2000}, (iii) gravito-inertial modes or Rossby (r) modes where the Coriolis force is the restoring force \citep{saio2018}, and (iv) stochastic solar-like oscillations for which pressure is the restoring force \citep[e.g.,][]{kjeldsen1995}. 
In addition, hybrid stars showing both $\delta$ Scuti and $\gamma$ Doradus pulsations have been detected and studied in detail in particular since the advent of space missions for high-precision photometry  \citep[e.g.,][]{grigahcene2010}. 

Cool stars typically rotate slowly, while hotter stars mostly show fast rotation. 
%The F-type star $\beta$\,Cas is an exception in this context as it rotates with $\sim$92\% of its critical velocity \citep{che2011}.
A high rotation rate leads to strong centrifugal forces which make a star oblate. Consequently, the stars' surface temperatures vary across the latitudes due to gravity darkening \citep{vonzeipel1924a,vonzeipel1924b}. Also, the distribution of the chemical elements in the stellar atmosphere is affected by high rotation rates \citep{meynet2000}. 

In massive and intermediate-mass O, B, and A stars on the main sequence, strong surface magnetic fields are found in $\sim$10\% of cases \citep[e.g.,][]{neiner2017}. These are fossil fields that can exist in stars down to effective temperatures of about 6500\,K \citep[e.g.,][]{2003A&A...404..669K}. They are usually dipolar with a typical strength of 3\,kG \citep{shultz2019}. At lower effective temperatures up to about 6700\,K, observed magnetic fields have a dynamo character \citep[e.g.,][]{2014MNRAS.444.3517M}. The exact transition between fossil and dynamo fields \citep{seach2020}, and the possible interaction between these two types of fields \citep{featherstone2009}, are not well known yet, but it is likely that some early F stars host both a fossil and a dynamo field observable at their surface. 

%subsection{The star $\beta$\,Cas}

$\beta$\,Cas (HD\,432, HR\,21, Caph) has an apparent magnitude of 2.27 in $V$ and a spectral type of F2\,III \citep{gray2003}. It is located at a heliocentric distance of 16.8\,$\pm$\,0.1\,pc \citep[determined from the Hipparcos parallax of 59.58\,$\pm$\,0.38\,mas;][]{vanLeeuwen2007}, and its mass is estimated to be 2.09\,\Msun\ \citep{holmberg2007}. $\beta$\,Cas is considered to be a rather evolved star located near the Terminal Age Main Sequence (TAMS) that was an A-type star during its main sequence lifetime.

Several measurements of $\beta$\,Cas' projected rotational velocity, \vsini, can be found in the literature: The first determination dates back to \citet{slettebak1955}, who reported a value of 70\,\kms. This is rather consistent with several recent measurements that lie between 69 and 71\,\kms\ \citep[e.g.,][]{glebocki2000,schroeder2009}.

A detailed interferometric study of $\beta$\,Cas \citep{che2011} yielded its geometric properties, surface temperature distribution, mass, and age. $\beta$\,Cas rotates with more than 90\% of its critical velocity, which causes significant radius and temperature differences between the poles and the equator: $\beta$\,Cas' radius is $\sim$24\% greater at the equator than at the poles and the temperature at the poles is $\sim$1000\,K higher than at the equator \citep{che2011}. The authors also determined the inclination angle to be  19.9\,$\pm$\,1.9$^{\circ}$, the rotation rate to be 1.12$^{+0.03}_{-0.04}$ d$^{-1}$, the mass to be 1.91\,$\pm$\,0.02\,\Msun, and the age to be 1.18\,$\pm$\,0.05\,Gyr.

\citet{millis1966} discovered the brightness variability of $\beta$\,Cas and identified it as a member of the class of $\delta$ Scuti stars as it showed a period of 0.104\,d (or 2.5\,hours) with an amplitude of 0.04\,mag in Johnson $V$. \citet{yang1982} confirmed the photometrically discovered period in radial velocity variations of $\beta$\,Cas, which had the highest amplitude in the Ca\,II line at 8662\,\ang.
\citet{antonello1986} refined the period to be 0.10101\,days with an amplitude of 0.03\,mag in $V$ based on eight nights of observations at Merate Observatory, and speculated that $\beta$\,Cas would be a small-amplitude, mono-periodic $\delta$ Scuti star.
$\delta$ Scuti pulsation and chromospheric variability were detected in the UV using data from the International Ultraviolet Explorer \citep[IUE;][]{ayres1991}; no hard evidence for a direct connection between both types of variations were found.
Based on Str\"omgren $uvby\beta$ observations, \citet{rodriguez1992} determined the pulsation frequency to be 9.91\,$\pm$\,0.35\,\cd\ (i.e., a pulsation period of 0.101\,d). The authors also determined the effective temperature, \Teff, of $\beta$\,Cas to be 7170\,K, its \logg\ to be 3.62 cgs, and its metallicity $[M/H]$ to be 0.2 from the Str\"omgren colours. Based on these findings, \citet{rodriguez1992} calculated the pulsation constant, $Q$, for the pulsation frequency to be 0.024 indicating first overtone radial pulsation, i.e., a p-mode with $n = 1$. These findings were reviewed by \citet{riboni1994} who confirmed the presence of a single pulsation frequency at 9.8997\,$\pm$\,0.0005\,\cd\,(i.e. a pulsation period of 0.10101\,d), and found \Teff\, of 7000\,$\pm$\,200\,K, \logg\ of 3.55\,$\pm$\,0.3 cgs, but also reported that a firm mode identification could not be performed.

$\beta$\,Cas' chromosphere was studied using IUE spectra by \citet{teays1989} who found that its chromospheric activity is modulated by the pulsation and that the mean level of chromospheric activity is comparable to other F-type stars.

%Binarity of beta Cas:
At the beginning of the 20$^{\rm th}$ century, it was speculated that $\beta$\,Cas could be a binary star with an orbital period of 27 days \citep{mellor1917}. However, a review of over 60 years of radial velocity data conducted by \citet{abt1965} revealed no sign of binarity.
A recent catalogue of $\delta$ Scuti stars in binary systems by \citet{liakos2017} lists $\beta$\,Cas as a `binary with an unspecified orbital period'.
%and bases its identification on the 2001 Washington Double Star Catalog \citep{mason2001} 

In this work we combine photometric time series obtained by the BRITE-Constellation \citep[where BRITE stands for BRIght Target Explorer;][]{weiss2014}, Solar Mass Ejection Imager \citep[SMEI;][]{eyles2003,jackson2004} and Transiting Exoplanet Survey Satellite \citep[TESS;][]{ricker2015} space telescopes to investigate the suggested mono-periodicity of $\beta$\,Cas. We also obtained spectropolarimetric observations with the Narval spectropolarimeter at the Telescope Bernard Lyot (TBL) and determine $\beta$\,Cas' magnetic properties. Based on these data, we also review the fundamental atmospheric parameters and chemical abundances of the star.

\section{Observations}
\subsection{BRITE-Constellation observations and data reduction}

The five 20x20x20-cm nanosatellites of the BRITE-Constellation\footnote{\url{http://www.brite-constellation.at}} each carry a 3-cm telescope feeding an uncooled CCD \citep{weiss2014}. 
Three BRITE satellites -- i.e., BRITE-Toronto (BTr), Uni-BRITE (UBr) and BRITE-Heweliusz (BHr) -- carry a custom-defined red filter (550 -- 700\,nm), and two satellites -- i.e., BRITE-Austria (BAb) and BRITE-Lem (BLb) -- a custom-defined blue filter (390 -- 460\,nm). More details on the detectors, pre-launch and in-orbit tests are described by \citet{pablo2016}. \citet{popowicz2017} describe the pipeline that processes the observed images yielding the instrumental magnitudes which are delivered to the users.

Each BRITE-Constellation nanosatellite observes large fields with typically 15 to 20 stars brighter than $V=6$ mag including at least three targets brighter than $V=3$ mag. Each field is observed  for at least 15 minutes each per $\sim$100-minute orbit for up to half a year. 

$\beta$\,Cas was observed during four consecutive seasons with the BRITE-Constellation nanosatellites: (i) in field 11-CasCep-I-2015 from 23 July to 1 November, 2015, using four of the five satellites: BRITE-Austria (BAb) and BRITE-Lem (BLb) with a blue filter, as well as BRITE-Heweliusz (BHr) and BRITE-Toronto (BTr) with a red filter, (ii) in field 19-Cas-I-2016 from 7 August, 2016, to 1 February, 2017, using the satellites BAb and UniBRITE (UBr, with a red filter), (iii) in field 30-Cas-II-2017 from 7 August, 2017, to 3 February, 2018 only using BAb, and (iv) in field 39-Cas-III-2018 from 7 August 2018 to 3 February 2019 using BAb and BTr. Table \ref{tab:obs} summarises the properties of the BRITE observations.

% New BRITE Data: 2018: BAb: 8.8.2018 to 14.11.2018, time base = 97.6d, N = 4443 (both values from the original files from Adam)
% BTr: 6.9.2018 - 18.10.2018 (setup 1) ,18.10.2018 - 21.1.2019 (setup 2); 41.3 d (setup 1), 14893 data points (setup 1); 95.4d (setup2) and 32521 data points (setup 2)
% Maybe expand the table and give also the dates of the observations and not write them explicitly in the text.

\begin{table*}[htb]
\caption{Properties of the BRITE-Constellation two-colour observations, the SMEI and TESS data for $\beta$\,Cas.}
\label{tab:obs}
\begin{center}
%\begin{scriptsize}
\begin{tabular}{llrrrrrrr}
\hline \hline 
 \multicolumn{1}{c}{Satellite} & \multicolumn{1}{c}{Field ID} & \multicolumn{1}{c}{Obs$_{\rm start}$} & \multicolumn{1}{c}{Obs$_{\rm end}$}& \multicolumn{1}{c}{Time base} & \multicolumn{1}{c}{1/T} & \multicolumn{1}{c}{N} & \multicolumn{1}{c}{Res. noise} & \multicolumn{1}{c}{f$_{\rm Nyq}$}  \\
\multicolumn{1}{c}{ } & \multicolumn{1}{c}{ }   & \multicolumn{1}{c}{ } & \multicolumn{1}{c}{ }   & \multicolumn{1}{c}{(d)} & \multicolumn{1}{c}{(\cd)} & \multicolumn{1}{c}{\#} & \multicolumn{1}{c}{(ppm)} & \multicolumn{1}{c}{(\cd)}  \\
\hline
BAb &	11-CasCep-I-2015	& 29 Aug. 2015  & 26 Oct. 2015  &   58.034	& 0.017	& 14305	& 163.1 & 2117.33 \\
BLb & 	11-CasCep-I-2015				& 29 Sep. 2015  & 17 Oct. 2015  & 18.067	& 0.055	& 8403	& 217.2 & 2119.62 \\
{\it BAb + BLb} & {\it combined} & {\it 29 Aug. 2015}   & {\it 26 Oct. 2015} &      {\it 58.034} & {\it 0.017} & {\it 22708} & {\it 130.8} & {\it 2133.35} \\
BHr & 	11-CasCep-I-2015				& 31 Aug. 2015  & 17 Oct. 2015  & 46.950	& 0.021	& 4552	& 215.4 & 2096.39 \\
BTr & 	11-CasCep-I-2015			& 4 Dec. 2015  & 20 Jan. 2016	& 43.681	& 0.023	& 20609	& 50.8 & 1892.47 \\
{\it BHr + BTr} & {\it combined} & {\it 31 Aug. 2015}	& {\it 20 Jan. 2016}  &	 {\it 142.096} &  {\it 0.007} & {\it 25161} & {\it 62.8} & {\it 1925.06} \\
\hline 
BAb & 19-Cas-I-2016		& 7 Aug. 2016  & 30 Dec. 2016  &   145.045	& 0.007	& 17728 	& 114.4 & 2085.80  \\
UBr & 	19-Cas-I-2016	& 13 Sep. 2016  & 1 Feb. 2017  &   141.072	& 0.007	& 23494 	& 84.2 & 2120.99 \\
\hline
BAb &	30-Cas-II-2017		& 7 Aug. 2017  & 3 Feb. 2018  & 160.170		&	0.006	&	7253	& 	227.2 & 2092.82	\\
\hline
BAb &     39-Cas-III-2018     &  8 Aug. 2018   & 14 Nov. 2018  & 96.587        &    0.010       &    1316       &      677.3   & 2075.96  \\
BTr &   39-Cas-III-2018      & 14 Sep. 2018 & 21 Jan. 2019  & 129.618        &    0.008       &   41757        &      92.4  & 2108.16   \\
\hline
SMEI	&	& 		&   &   			2884.825 & 0.0003 & 28742 & 64.8  & 7.08 \\
\hline
TESS &    Sector 17     &  7 Oct. 2019  & 2 Nov. 2019  &    23.081      &      0.043     &     13134       &     -  & -     \\
  &    Sector 18     &  2 Nov. 2019  & 27 Nov. 2019  &   22.873       &     0.044       &      14961      &     -   & -   \\
  &    {\it combined}     &  {\it 7. Oct. 2019}  & {\it 27 Nov. 2019}  & {\it 45.954}        &     {\it 0.02}      &      {\it 28095}     &    {\it 2.4}    & {\it 359.39}   \\
\hline
\end{tabular}
%\end{scriptsize}
\end{center}
\tablefoot{Satellite used to conduct the observations (Satellite), BRITE-Constellation field ID and TESS Sector number (Field ID), corresponding start (Obs$_{\rm start}$) and end dates (Obs$_{\rm end}$) of observations, total time base of the reduced data set (Time base), Rayleigh frequency resolution (1/T), number of data points (N), residual noise after prewhitening all frequencies (Res. noise) which is calculated over the complete frequency range relevant for $\delta$ Scuti pulsations from 0 to 100\,\cd, and the Nyquist frequency (f$_{\rm Nyq}$) for each data set.  As the TESS data of sectors 17 and 18 were analysed together, no individual values for the residual noise and Nyquist frequency are provided.}
\end{table*}

\begin{figure*}
\begin{center}
\includegraphics[width=0.9\textwidth]{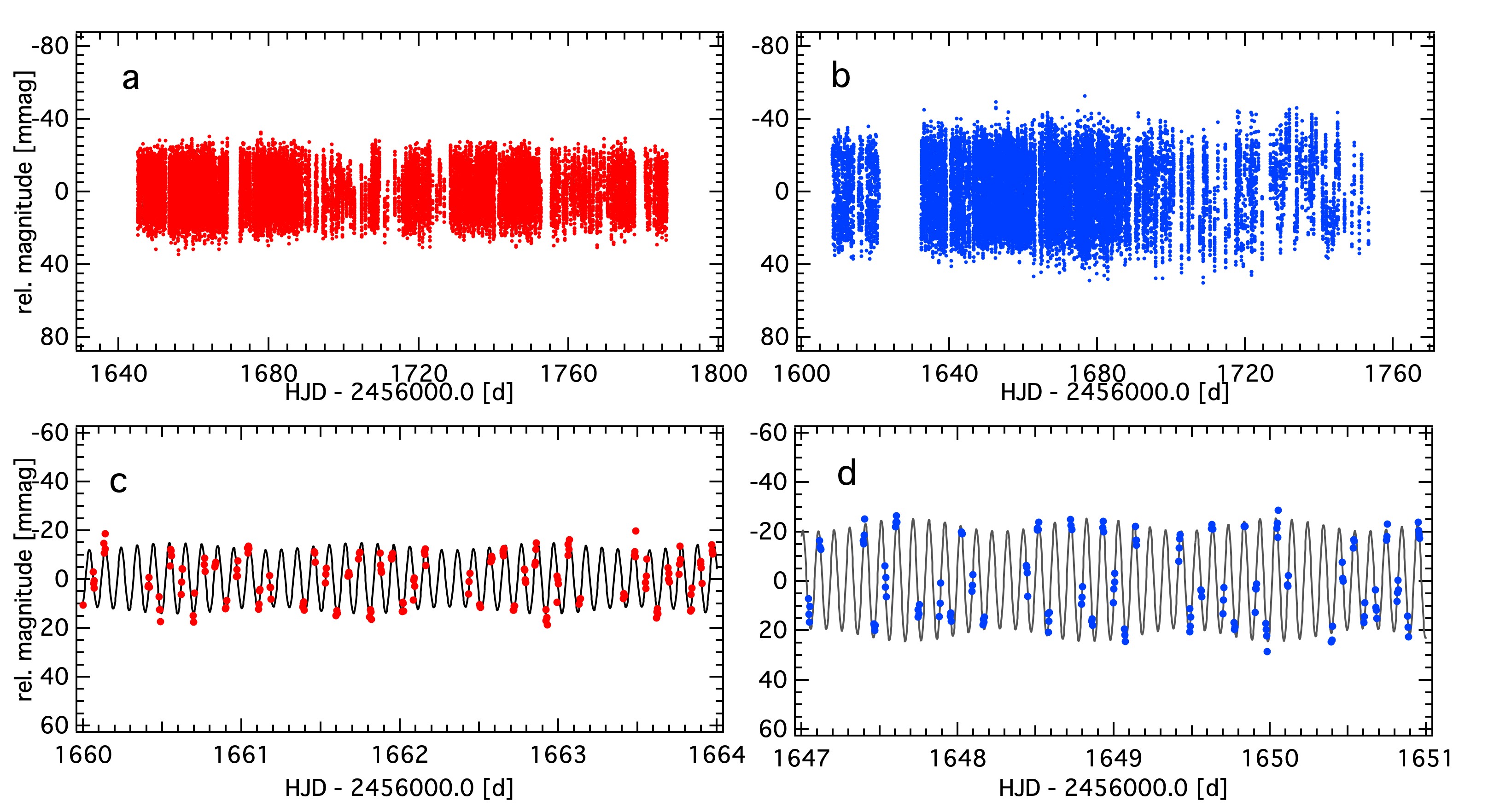}
\caption[]{BRITE photometric time series obtained by UBr (panel a) and BAb in 2016 (panel b) to the same Y-axis scale and with a time base of 170\,days on both X axes. Panels c and d show 4-days subsets of the UBr and BAb 2016 light curves binned to 5-minute intervals and the corresponding multi-sine fit with the two identified pulsation frequencies again to the same Y-axis scale and with a time base of 4\,days on both X axes.}
\label{fig:BRITE2016-lcs}
\end{center}
\end{figure*}

The BRITE raw data were extracted from the 2D images following the procedure described by \cite{popowicz2017}. The BRITE photometry was subsequently corrected for instrumental effects. The corrections included outlier rejection, and both one- and two-dimensional decorrelations with all available parameters, in accordance with the procedure described by \citet{pigulski2018}.
Figure~\ref{fig:BRITE2016-lcs} shows the full light curves obtained by UBr (panel a) and BAb (panel b) in 2016 as well as 4-day subsets of the data illustrating the pulsational variability. The corresponding light curves of the three other seasons are given in the Appendix (Fig.~\ref{fig:BRITE-lcs-appendix}).

\subsection{Solar Mass Ejection Imager (SMEI) observations and data reduction}

The SMEI experiment \citep{eyles2003,jackson2004}, placed on-board the Coriolis spacecraft, aimed at measuring sunlight scattered by free electrons of the solar wind, but the images are also suitable to extract the photometry of bright stars. The $\beta$\,Cas data were obtained in the years 2003\,--\,2010 and are  available through the University of California San Diego (UCSD) web page\footnote{\url{http://smei.ucsd.edu/new\_smei/index.html}}. The SMEI photometry is affected by long-term calibration effects, especially a spurious variability with a period of one year. We corrected the SMEI UCSD photometry of $\beta$\,Cas for this one-year variability by subtracting an interpolated mean light curve. 
%In addition, the worst parts of the light curve {\color{red} WHICH PART?} and outliers were removed. 
Finally, the individual uncertainties were calculated for each data point using the scatter of the neighbouring data points, and the worst parts of the light curve (i.e., those with an uncertainty higher than 0.013\,mag) and outliers were removed. The low-frequency instrumental variability was filtered out by subtracting detrended residuals of the fit. This procedure removed the intrinsic low-frequency (frequencies below $\sim$1~d$^{-1}$) variability, if present.

The final SMEI data for $\beta$\,Cas comprise 28742 data points obtained between 6 February 2003 and 31 December 2010 for a total time base of 2884.89537\,d ($\sim$7.9\,yr). This corresponds to a Rayleigh frequency resolution, $1/T$, of 0.0003\cd. An overview of the properties of the SMEI data is given in Table \ref{tab:obs}.

\subsection{TESS data}

The TESS satellite \citep{ricker2015} was launched on 18 April 2018 on-board a Falcon 9 rocket and carries four identical cameras. Each camera has an effective aperture size of 10\,cm and a $24^{\circ} \times 24^{\circ}$ wide field of view. The red-optical TESS bandpass ranges from 600 to 1000\,nm. TESS observations are conducted almost across the entire sky in 13 sectors each with a time base of $\sim$27\,days. Depending on the position of a given target in the sky, TESS data can span between a minimum of $\sim$\,27 and a maximum of 351 days. TESS photometric time series are typically conducted at a cadence of 30\,minutes. For a sub-sample of 200\,000 stars, observations are carried out at a 2-minute cadence. TESS data are made available to the public through the "Barbara A. Mikulski Archive for Space Telescopes'' (MAST)\footnote{\url{http://archive.stsci.edu}}.

TESS's primary mission is the detection of small exoplanets transiting bright, nearby stars. As a secondary science goal, the TESS data is used for asteroseismic studies of all classes of pulsating stars across the Hertzsprung-Russell (HR) diagram. 

We combined the photometric time series obtained from BRITE-Constellation and SMEI with observations taken by TESS from 7 October to 27 November 2019 in Sectors 17 and 18. The total time base of the TESS data for $\beta$\,Cas (TIC 396298498, TESS magnitude $T$ = 2.041 mag) is 45.954 days (see Table \ref{tab:obs}). 
$\beta$\,Cas is one of the pre-selected targets for 2-minute cadence observations.

\begin{figure*}
\begin{center}
\includegraphics[width=0.9\textwidth]{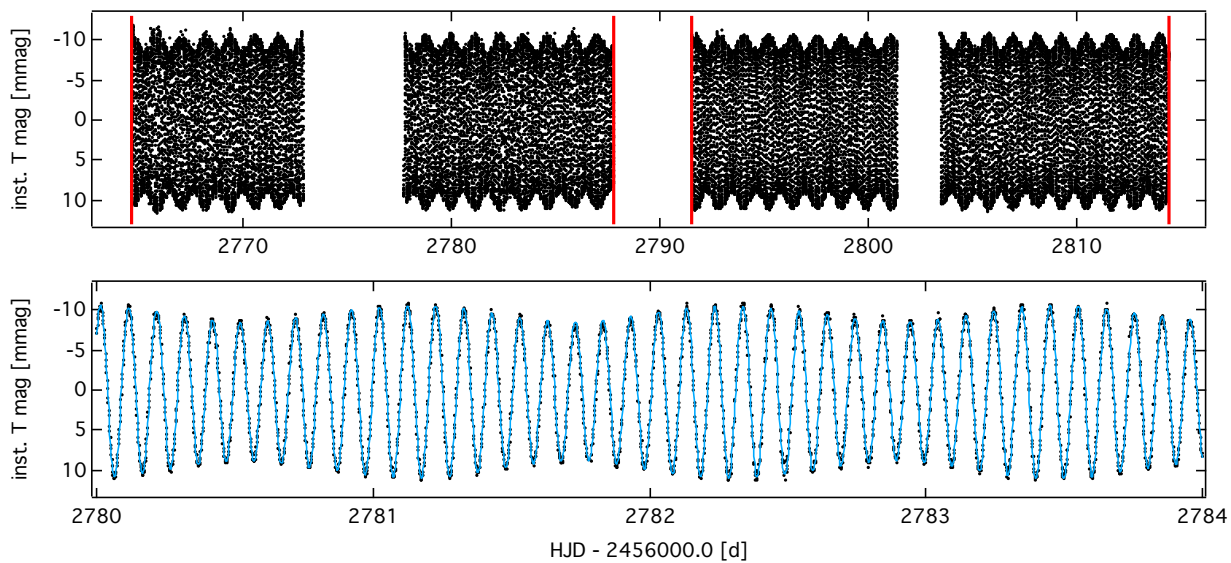}
\caption[]{Top panel: Complete TESS light curve . Start and end dates of Sectors 17 and 18 are marked with vertical red lines. Bottom panel: 4-day zoom illustrating the pulsational variability. }
\label{fig:TESS-lcs}
\end{center}
\end{figure*}

The full TESS light curve and a zoom into a 4-day subset are given in Fig.~\ref{fig:TESS-lcs}. The beginnings and ends of the two Sectors are marked with a vertical dashed line. The obvious gaps within each Sector are caused by the data downlink at the perigee of the TESS orbit. $\beta$\,Cas was observed with CCD 1 of Camera 2 in Sector 17  and with CCD 2 of Camera 2 in Sector 18.
%\footnote{\url{https://archive.stsci.edu/missions/tess/doc/tess_drn/tess_sector_18_drn25_v02.pdf}}. 
Just before the end of the first orbit, the Moon entered Camera 1 and 2's field-of-view. No other instrumental effects were reported (see the TESS Data Release Notes for Sector 17\footnote{{\url{https://archive.stsci.edu/missions/tess/doc/tess_drn/tess_sector_17_drn24_v02.pdf}}} and 18 \footnote{\url{https://archive.stsci.edu/missions/tess/doc/tess_drn/tess_sector_18_drn25_v02.pdf}}). 

We used the 2-minute Simple Aperture Photometry (SAP) flux light curve, provided by the MAST archive
%\footnote{http://archive.stsci.edu} 
and the Python packages \texttt{lightkurve} \citep{lightkurve} and \textit{SMURFS}\footnote{\url{https://github.com/MarcoMuellner/SMURFS}} \citep{muellner2020}. We removed all measurements with a non-zero "quality'' flag (see §9 in the TESS Science Data Products Description Document\footnote{\url{https://archive.stsci.edu/missions/tess/doc/EXP-TESS-ARC-ICD-TM-0014.pdf}}), which marks anomalies like cosmic rays, instrumental issues or stray light from the Earth or Moon. 
Furthermore, the outliers were removed by means of 4$\sigma$-clipping.
%Furthermore the light curve was sigma clipped, removing outliers outside of four $\sigma$. 
In the next step, we subtracted the median flux of the two sectors and combined them. At this point, a first frequency analysis was conducted to check for the presence of a low frequency signal that might originate from g-modes and of the rotational frequency of 1.12\,\cd\ as found by \citet{che2011}. Figure \ref{fig:TESS-zoom} shows a zoom into the low-frequency domain of the TESS data at that point in the analysis and illustrates that the rotation frequency cannot be detected. It is also evident that there is an increased noise level in the region between 0.0 and 0.5\,\cd\ that is of instrumental origin. No clear evidence for significant g-mode frequencies can be found. Hence, in the next step, 
%As no signs of {\bf the rotation frequency and } g-mode pulsations were detected,
we applied a Gaussian filter using \texttt{scipy} \citep{scipy} to remove the low frequency signal that is of instrumental origin.
%and subtracted it from the lightcurve. 
This procedure removed any intrinsic low-frequency (i.e., frequencies below $\sim$2~d$^{-1}$) variability, if present.

\subsection{Narval spectropolarimetric observations}
\label{sect:specpol}

$\beta$\,Cas was observed with the Narval spectropolarimeter installed at the 2.03-m T\'elescope Bernard Lyot (TBL) at Pic du Midi Observatory in France. Narval provides spectra covering the wavelength range from 3900 to 10500\,\AA with a resolving power of $\sim$65000, spread over 40 echelle orders recorded on a single detector. We used Narval in circular polarisation mode to produce Stokes V and Stokes I spectra from a sequence of 4 sub-exposures. A null polarisation spectrum (N) is also produced by combining the 4 sub-exposures in a destructive way. N allows to check for signal due to instrumental effects or stellar phenomena unrelated to magnetism, such as pulsations.

$\beta$\,Cas was observed a first time on November 3, 2013, as a part of the BRITEpol survey \citep{neiner2016}. BRITEpol measures the potential magnetic field of all stars brighter than $V=4$~mag, as a ground-based support to BRITE. The possible detection of a magnetic field in this first observation of $\beta$\,Cas led to a series of three additional observations obtained between September 24 and December 21, 2014, to confirm the presence of a magnetic field in $\beta$\,Cas. Finally, a complete series of follow-up observations was acquired between December 1 and 13, 2015, simultaneously with the BRITE observations.

Narval employs the beam-exchange polarimetric modulation procedure \citep{semel1993,donati1997,bagnulo2009}, which allows one to compensate most instrumental polarisation artefacts and prevent a temporal variation of stellar spectra to result in spurious polarisation signatures. In any case, due to brightness of the star, the individual sub-exposures could be kept very short (65~s), excluding any possibility of spurious polarisation appearing due to spectral variability associated with $\delta$~Scuti pulsations.
To obtain a higher signal-to-noise (S/N), several consecutive polarimetric sequences were then recorded and co-added after reduction (see Sect.~\ref{sect:LSD}).

The data were reduced with the {\sc Libre-Esprit} software pipeline available at TBL. Each order of the intensity spectra was then normalised separately using {\sc SPENT} \citep{martin2018} and the normalisation function was also applied to Stokes V and N. 

The log of the Narval observations is available in Table~\ref{logNarval}.

\begin{table}
\caption{Journal of the spectropolarimetric observations.}
\begin{tabular}{lllll}
\hline \hline 
\# & Date & mid-HJD & \multicolumn{1}{c}{$T_{\rm exp}$} & S/N \\
   &      & -2450000 & \multicolumn{1}{c}{[s]} & \\
\hline
1  & Nov 3, 2013  & 6600.2627 & \,\,\,\,\,\,\,4$\times$65 & 818\\
2  & Sep 25, 2014 & 6925.6165 & 3$\times$4$\times$65 & 1206\\
3  & Dec 19, 2014 & 7011.2941 & 5$\times$4$\times$65 & 1258\\
4  & Dec 21, 2014 & 7013.2990 & 5$\times$4$\times$65 & 1122\\
5  & Dec 1, 2015  & 7358.3265 & 5$\times$4$\times$65 & 984\\
6  & Dec 1, 2015  & 7358.3891 & 5$\times$4$\times$65 & 881\\
7  & Dec 2, 2015  & 7359.3314 & 5$\times$4$\times$65 & 723\\
8  & Dec 2, 2015  & 7359.3870 & 5$\times$4$\times$65 & 781\\
9  & Dec 5, 2015  & 7362.2862 & 5$\times$4$\times$65 & 918\\
10  & Dec 6, 2015 & 7363.2947 & 5$\times$4$\times$65 & 1100\\
11  & Dec 6, 2015 & 7363.3535 & 5$\times$4$\times$65 & 1154\\
12  & Dec 6, 2015 & 7363.4058 & 5$\times$4$\times$65 & 1160\\
13  & Dec 7, 2015 & 7364.3514 & 5$\times$4$\times$65 & 1140\\
14  & Dec 7, 2015 & 7364.4070 & 5$\times$4$\times$65 & 1058\\
15 & Dec 9, 2015  & 7366.3073 & 5$\times$4$\times$65 & 1040\\
16 & Dec 11, 2015 & 7368.3693 & 5$\times$4$\times$65 & 1093\\
17 & Dec 12, 2015 & 7369.3194 & 5$\times$4$\times$65 & 991\\
18 & Dec 12, 2015 & 7369.3734 & 5$\times$4$\times$65 & 889\\
19 & Dec 13, 2015 & 7370.3799 & 5$\times$4$\times$65 & 995\\
\hline
\end{tabular}
\label{logNarval}
\tablefoot{Indicated are the index number of the averaged polarimetric measurement, the date of observations, the Heliocentric Julian Date at the middle of the observations (mid-HJD - 2450000), the number of polarimetric sequences times the exposure time in seconds, and the average signal-to-noise ratio of a (single) spectropolarimetric sequence per CCD pixel at $\sim$500 nm.}
\end{table}

\section{Photometric analysis}

\begin{table*}
\caption{Pulsation frequencies (F), amplitudes ({\it A}), phases ($\phi$), and signal-to-noise values (S/N) of $\beta$\,Cas derived from the BRITE-Constellation, SMEI and TESS data.}
\label{tab:freqs}
\begin{center}
%\begin{scriptsize}
\begin{tabular}{rrrrrrrr}
\hline
\hline
\multicolumn{1}{c}{ } & \multicolumn{1}{c}{F$_1$} & \multicolumn{1}{c}{F$_2$} & \multicolumn{1}{c}{F$_3$ = 2$\cdot$F$_1$} & \multicolumn{1}{c}{F$_4$} & \multicolumn{1}{c}{F$_5$ = $F_1 + F_2$} & \multicolumn{1}{c}{F$_6$ = $F_1\pm(1/T)$} & \multicolumn{1}{c}{ } \\
\hline
Freq. {\it BRITE red}	&	9.89708(1)	&	9.0437(1)	&	19.7922(6)	&	\multicolumn{1}{c}{-}	&	\multicolumn{1}{c}{-}	&	\multicolumn{1}{c}{-}	&	\cd	\\
$A_{R}$ 2015	&	13.30(4)	&	1.51(4)	&	0.28(4)	&	\multicolumn{1}{c}{-}	&	\multicolumn{1}{c}{-}	&	\multicolumn{1}{c}{-}	&	mmag	\\
$A_{R}$ 2016	&	13.14(6)	&	1.43(6)	&	0.40(6)	&   \multicolumn{1}{c}{-}	&	\multicolumn{1}{c}{-}	&	\multicolumn{1}{c}{-}	&	mmag	\\
$A_{R}$ 2018	&	12.57(6)	&	1.34(6)	&	\multicolumn{1}{c}{-}	&	\multicolumn{1}{c}{-}	&	\multicolumn{1}{c}{-}	&	\multicolumn{1}{c}{-}	&	mmag	\\
$\phi_R$ 2015	&	0.1911(5)	&	0.216(5)	&	0.4(2)	&	\multicolumn{1}{c}{-}	&	\multicolumn{1}{c}{-}	&	\multicolumn{1}{c}{-}	&	 	\\
$\phi_R$ 2016	&	0.1505(7)	&	0.769(6)	&	0.4(2)	&	\multicolumn{1}{c}{-}	&	\multicolumn{1}{c}{-}	&	\multicolumn{1}{c}{-}	&	 	\\
$\phi_R$ 2018	&	0.1566(7)	&	0.791(6)	&	\multicolumn{1}{c}{-}	&	\multicolumn{1}{c}{-}	&	\multicolumn{1}{c}{-}	&	\multicolumn{1}{c}{-}	&	 	\\
S/N$_R$ 2015	&	21.47	&	21.94	&	4.38	&	\multicolumn{1}{c}{-}	&	\multicolumn{1}{c}{-}	&	\multicolumn{1}{c}{-}	&		\\
S/N$_R$ 2016	&	23.81	&	17.57	&	4.98	&	\multicolumn{1}{c}{-}	&	\multicolumn{1}{c}{-}	&	\multicolumn{1}{c}{-}	&		\\
S/N$_R$ 2018	&	32.54	&	14.46	&	\multicolumn{1}{c}{-}	&	\multicolumn{1}{c}{-}	&	\multicolumn{1}{c}{-}	&	\multicolumn{1}{c}{-}	&		\\
\hline
Freq. {\it BRITE blue}	&	9.89710(1)	&	9.0434(1)	&	19.7946(7)	&	\multicolumn{1}{c}{-}	&	\multicolumn{1}{c}{-}	&	\multicolumn{1}{c}{-}	&	\cd	\\
$A_{B}$ 2015	&	22.2(1)	&	2.5(1)	&	0.5(1)	&	\multicolumn{1}{c}{-}	&	\multicolumn{1}{c}{-}	&	\multicolumn{1}{c}{-}	&	mmag	\\
$A_{B}$ 2016	&	22.47(8)	&	2.62(8)	&	0.47(8)	&	\multicolumn{1}{c}{-}	&	\multicolumn{1}{c}{-}	&	\multicolumn{1}{c}{-}	&	mmag	\\
$A_{B}$ 2017	&	22.2(2)	&	2.38(2)	&	\multicolumn{1}{c}{-}	&	\multicolumn{1}{c}{-}	&	\multicolumn{1}{c}{-}	&	\multicolumn{1}{c}{-}	&	mmag	\\
$A_{B}$ 2018	&	23.7(4)	&	\multicolumn{1}{c}{-}	&	\multicolumn{1}{c}{-}	&	\multicolumn{1}{c}{-}	&	\multicolumn{1}{c}{-}	&	\multicolumn{1}{c}{-}	&	mmag	\\
$\phi_B$ 2015	&	0.1719(7)	&	0.132(7)	&	0.49(3)	&	\multicolumn{1}{c}{-}	&	\multicolumn{1}{c}{-}	&	\multicolumn{1}{c}{-}	&	 	\\
$\phi_B$ 2016	&	0.1469(6)	&	0.82(5)	&	0.99(3)	&	\multicolumn{1}{c}{-}	&	\multicolumn{1}{c}{-}	&	\multicolumn{1}{c}{-}	&	 	\\
$\phi_B$ 2017	&	0.126(1)	&	0.78(1)	&	\multicolumn{1}{c}{-}	&	\multicolumn{1}{c}{-}	&	\multicolumn{1}{c}{-}	&	\multicolumn{1}{c}{-}	&	 	\\
$\phi_B$ 2018	&	0.625(3)	&	\multicolumn{1}{c}{-}	&	\multicolumn{1}{c}{-}	&	\multicolumn{1}{c}{-}	&	\multicolumn{1}{c}{-}	&	\multicolumn{1}{c}{-}	&	 	\\
S/N$_B$ 2015	&	17.41	&	18.67	&	3.53	&	\multicolumn{1}{c}{-}	&	\multicolumn{1}{c}{-}	&	\multicolumn{1}{c}{-}	&		\\
S/N$_B$ 2016	&	23.15	&	21.45	&	3.99	&	\multicolumn{1}{c}{-}	&	\multicolumn{1}{c}{-}	&	\multicolumn{1}{c}{-}	&		\\
S/N$_B$ 2017	&	14.12	&	10.27	&	\multicolumn{1}{c}{-}	&	\multicolumn{1}{c}{-}	&	\multicolumn{1}{c}{-}	&	\multicolumn{1}{c}{-}	&		\\
S/N$_B$ 2018	&	7.49	&	\multicolumn{1}{c}{-}	&	\multicolumn{1}{c}{-}	&	\multicolumn{1}{c}{-}	&	\multicolumn{1}{c}{-}	&	\multicolumn{1}{c}{-}	&		\\
\hline
Freq.$_{SMEI}$	&	9.8971699(9)	&	9.044955(8)	&	\multicolumn{1}{c}{-}	&	\multicolumn{1}{c}{-}	&	\multicolumn{1}{c}{-}	&	\multicolumn{1}{c}{-}	&	\cd	\\
$A_{SMEI}$	&	11.25(5)	&	1.24(5)	&	\multicolumn{1}{c}{-}	&	\multicolumn{1}{c}{-} &	\multicolumn{1}{c}{-}	&	\multicolumn{1}{c}{-}	&	mmag	\\
$\phi_{SMEI}$	&	0.1301(7)	&	0.001(7)	&	\multicolumn{1}{c}{-}	&	\multicolumn{1}{c}{-}	&	\multicolumn{1}{c}{-}	&	\multicolumn{1}{c}{-}	&	 	\\
S/N$_{SMEI}$	&	120.65	&	21.41	&	\multicolumn{1}{c}{-}	&	\multicolumn{1}{c}{-}	&	\multicolumn{1}{c}{-}	&	\multicolumn{1}{c}{-}	&		\\
\hline
Freq.$_{TESS}$	&	9.897098(2)	&	9.04391(2)	&	19.7942(1)	&	8.3847(4)	&	18.9409(6)	&	9.9000(5)	&	\cd	\\
$A_{TESS}$	&	9.749(2)	&	1.035(2)	&	0.229(2)	&	0.055(2)	&	0.038(2)	&	0.025(2)	&	mmag	\\
$\phi_{TESS}$	&	0.77974(4)	&	0.4214(3)	&	0.304(2)	&	0.771(6)	&	0.260(9)	&	0.305(8)	&	 	\\
S/N$_{TESS}$	&	20.28	&	153.52	&	54.77	&	10.51	&	10.65	&	19.365	&		\\\hline
\end{tabular}
%\end{scriptsize}
\end{center}
\tablefoot{Frequency, amplitude and phase errors are given as last-digit errors in parentheses and were calculated following \citet{montgomery99}. Frequencies F$_3$, F$_5$, and F$_6$ are identified as linear combinations in the top line. }
\end{table*}

The frequency analysis of the BRITE, SMEI and TESS photometric time series of $\beta$\,Cas was performed independently of each other using the software package Period04 \citep{lenz05} that combines Fourier and least-squares algorithms. Frequencies were then prewhitened and considered to be significant if their amplitudes exceeded 3.9 times the local noise level in the amplitude spectrum \citep{breger93,kuschnig97}.
Frequency, amplitude and phase errors are calculated using the formulae given by \citet{montgomery99}.
To verify the analysis, we use the frequency extraction tool \textit{SMURFS}, which automates the search for significant frequencies in time series data by iterative searching for the frequency with the maximum amplitude and removing it using curve fitting tools provided by \texttt{lmfit} \citep{lmfit} and \texttt{scipy} \citep{scipy}.

%$\beta$\,Cas shows three frequencies that can be attributed to \dsct type pulsations: F$_1$ at 9.89708\,\cd\ and F$_2$ at 9.04369\,\cd\ clearly dominate and can be identified in all data sets. The third independent \dsct type frequency is F$_4$ at 8.3847\,\cd, which is detected only in the TESS data due to its relatively small amplitude of 0.055\,mmag.

All analysed data sets show the presence of two pulsation modes with frequencies F$_1$ at 9.89708\,\cd\ and F$_2$ at 9.04369\,\cd\ (see Table \ref{tab:freqs}). In some of the BRITE data sets and the TESS data, an harmonic frequency to $F_1$ was detected. Only in the TESS data a third independent \dsct type frequency of F$_4$ at 8.3847\,\cd\ with an amplitude of only 0.055\,mmag and a combination frequency of F$_1$ + F$_2$ were found.

The harmonic frequency F$_3$ at 19.79221\,\cd, is identified to be two times F$_1$. It is only significant in those photometric time series that have a low enough noise level (i.e., in the BRITE 2015 and 2016 B \& R observations and in TESS data). 
Similarly, frequency F$_5$ at 18.9409\,\cd\ is actually F$_1$ plus F$_2$; due to its small amplitude of 0.038\,mmag it is only detectable in the TESS data. The TESS data also allow us to identify a frequency F$_6$ at 9.9000\,\cd\ that is the same as F$_1$ within the Rayleigh frequency resolution. We speculate that it results from the small amplitude changes detected in F$_1$ (see Sect. \ref{sect:ampvar}). With the presently available data, we cannot decide unambiguously whether F$_6$ is an independent frequency or is dependent on F$_1$. Future additional TESS observations of $\beta$\,Cas will hopefully contribute to finding an unambiguous solution.

A summary of the pulsation frequencies, amplitudes and phases derived from the three seasons of BRITE-Constellation data, SMEI and TESS observations is given in Table \ref{tab:freqs}. 

\begin{figure*}
\begin{center}
\includegraphics[width=0.9\textwidth]{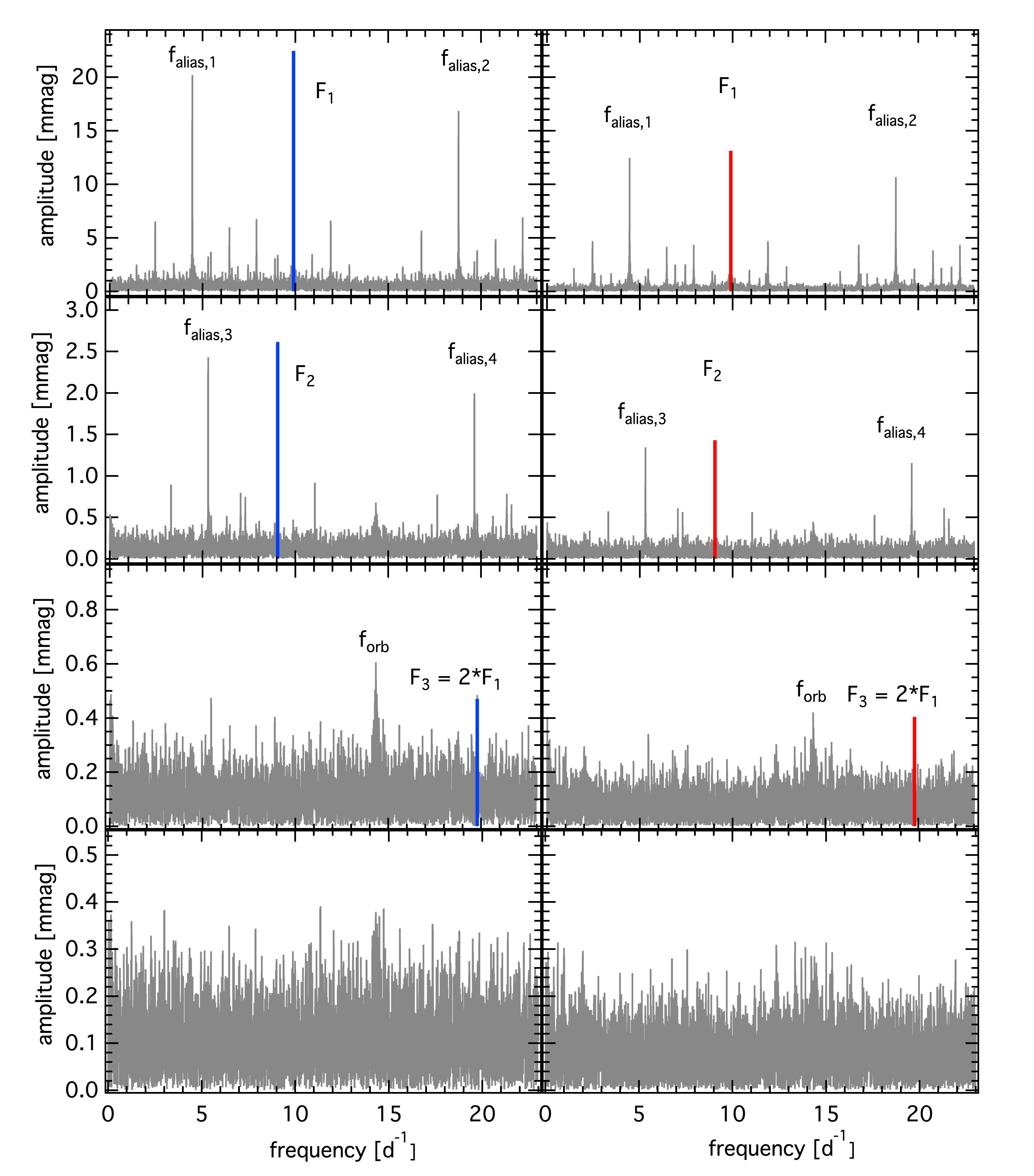}
\caption[]{Amplitude spectra of the BRITE-Constellation data obtained in 2016: BAb blue filter data are shown on the left side, UBr red filter data on the right side. Top panels show the amplitude spectra of the original data with F$_1$ identified, the second panels illustrate the amplitude spectra after prewhitening F$_1$, the third panels those after prewhitening F$_1$ and F$_2$, and the bottom panel displays the residual amplitude spectra after prewhitening all significant frequencies. An explanation for the identified alias frequencies is given in the text.}
\label{amps2016}
\end{center}
\end{figure*}

As an example, the amplitude spectra from the 2016 BAb and UBr data are shown in Fig.~\ref{amps2016}: blue filter data are shown on the left, red filter data on the right side. The top panels illustrate the amplitude spectra of the original data with F$_1$ identified as well as two alias frequencies, which appear in the blue and the red filter data. f$_{alias,1}$ can be identified as the corresponding BRITE orbital frequency, f$_{orb}$, minus F1 and f$_{alias,2}$ as 2 times f$_{orb}$ minus F1. Both alias frequencies disappear after prewhitening F$_1$ as can be seen in the middle panels of Fig.~\ref{amps2016}, which shows the amplitude spectra after subtraction of F$_1$, where F$_2$ and two new alias frequencies f$_{alias,3}$ and f$_{alias,4}$ are marked. f$_{alias,3}$ is generated as f$_{orb}$ minus F2, and f$_{alias,4}$ as 2 times f$_{orb}$ minus F2. They again disappear when F$_2$ is prewhitened. The bottom panels show the amplitude spectra after prewhitening with F$_1$ and F$_2$ with F$_3 = 2 \cdot$F$_1$ and f$_{orb}$ marked. The residual amplitude spectra for the 2016 blue and red filter data are shown in the bottom panel of Fig.~\ref{amps2016} and illustrate the corresponding average residual noise levels of 114.4\,ppm for BAb and 84.2\,ppm for UBr.
The amplitude spectra for the 2015, 2017, and 2018 BRITE observations, and the spectral window functions are given in Appendix \ref{app} (Figs. \ref{amps2015}, \ref{amps2017}, \ref{amps2018} , and \ref{BRITE-spws}).

%\begin{figure}
%\begin{center}
%\includegraphics[width=0.49\textwidth]{betaCas2016_residuals.jpg}
%\caption[]{Residual amplitude spectra for the 2016 BAb data (in blue pointing upwards) and UBr data (in red pointing downwards) after prewhitening all significant frequencies.}
%\label{residuals2016}
%\end{center}
%\end{figure}

\begin{figure}
\begin{center}
\includegraphics[width=0.47\textwidth]{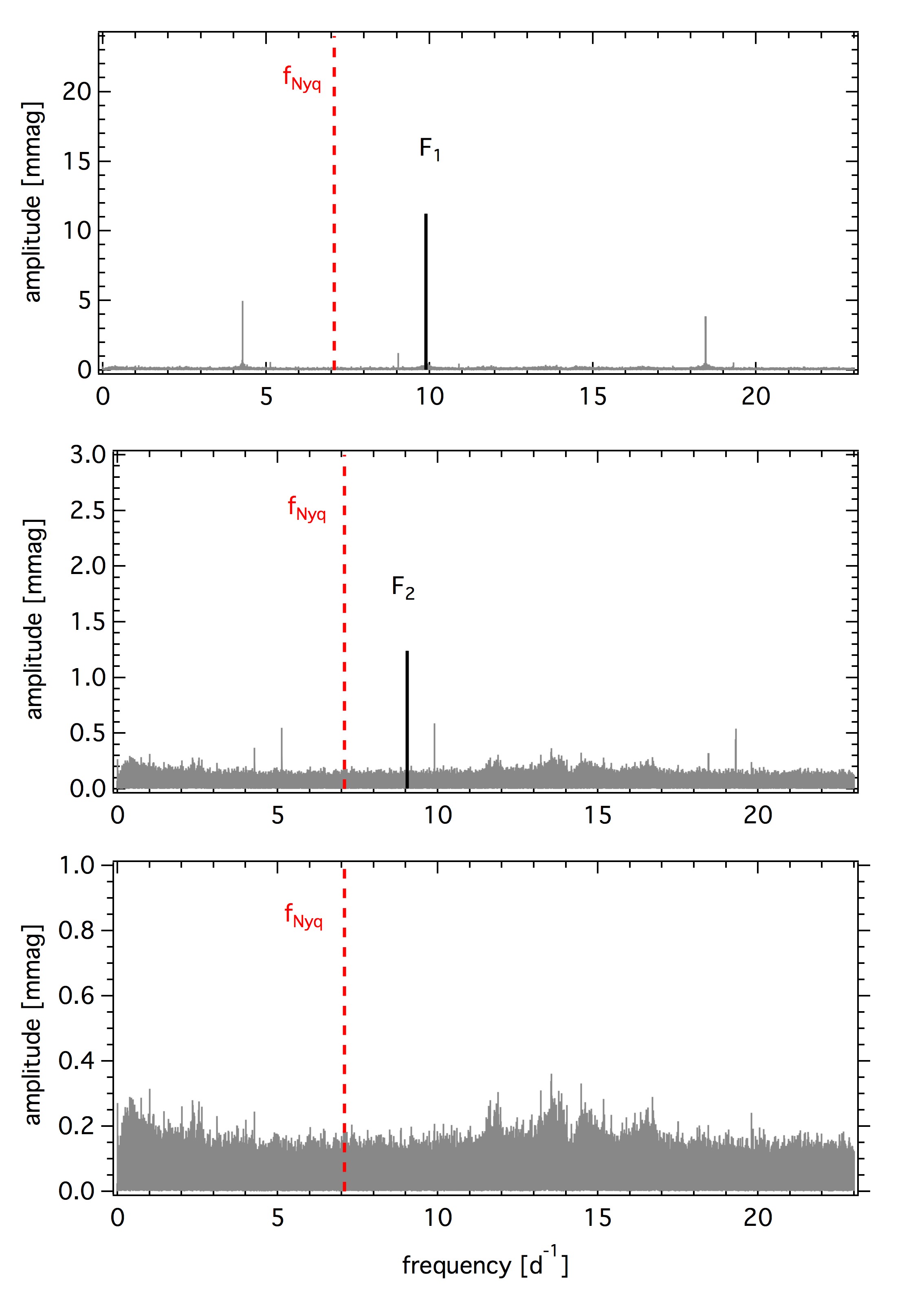}
\caption[]{Amplitude spectra of the SMEI data: original data with F$_1$ identified (top panel), amplitude spectra after prewhitening with F$_1$ and showing F$_2$ (middle panel), and residuals after prewhitening F$_1$ and F$_2$ (bottom panel). The position of the Nyquist frequency is marked in red.}
\label{SMEI_ampspecs}
\end{center}
\end{figure}

The Nyquist frequency for the SMEI data is at 7.08\,\cd\ because measurements were taken every $\sim$1.7\,hours. The frequencies of \dsct stars are typically higher than this value \citep[e.g.,][]{aerts2010}. As it was shown for Kepler data by \citet{murphy2013}, it is possible to do super-Nyquist asteroseismology using the SMEI data because the real peaks remain as singlets even if they are above $f_{\rm Nyq}$. Using the SMEI data, we confirm the pulsation frequencies F$_1$ and F$_2$ (see Fig.~\ref{SMEI_ampspecs}). The residual amplitude spectrum after prewhitening F$_1$ and F$_2$ is shown in the bottom panel of Fig.~\ref{SMEI_ampspecs}. Figure~\ref{SMEI-spw} shows the SMEI spectral window.

\begin{figure}
\begin{center}
\includegraphics[width=0.49\textwidth]{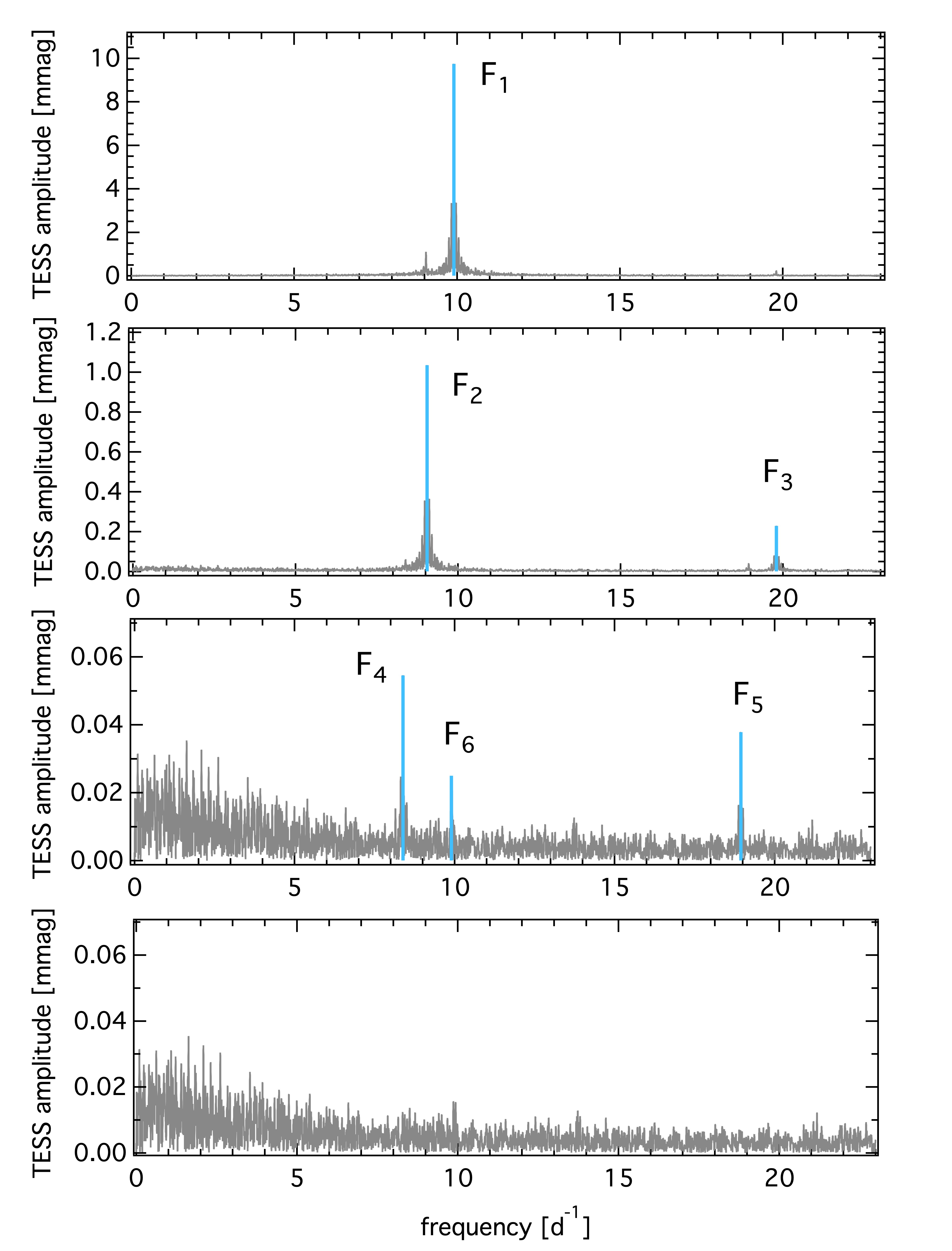}
\caption[]{Fourier analysis of the TESS data: The amplitude spectrum of the original data with F$_1$ identified ( top panel), amplitude spectrum after prewhitening F$_1$ (second panel from the top), amplitude spectrum after prewhitening F$_1$, F$_2$, and F$_3$ (third panel from the top), and residual amplitude spectrum after prewhitening all significant frequencies (bottom panel).}
\label{fig:tess-amps}
\end{center}
\end{figure}

The amplitude spectra obtained from the TESS data are illustrated in Fig.~\ref{fig:tess-amps}: following the prewhitening sequence, F$_1$ to F$_6$ are identified and marked, respectively. The residual amplitude spectrum after prewhitening the six frequencies is shown in the bottom panel of Fig.~\ref{fig:tess-amps}. For completeness, Fig.~\ref{TESS-spw} illustrates the spectral window of the TESS data.

%\begin{figure}
%\begin{center}
%\includegraphics[width=0.5\textwidth]{TESS-residuals.jpg}
%\caption[]{Residual amplitude spectrum of the TESS data after prewhitening all significant frequencies.}
%\label{fig:tess-residuals}
%\end{center}
%\end{figure}

\section{Spectroscopic analysis}
\label{Sec:spectro}

\subsection{Atmospheric parameters and chemical abundances}
% Part by Tanya

The spectropolarimetric time-series were averaged to produce one spectrum with $S/N=1250$ estimated in a small region with only continuum from 6070--6071 \AA. This value for the signal-to-noise ratio corresponds to the actual pixel-to-pixel scatter measured in the red. It is not much larger than the values given in Section \ref{sect:specpol} due to a fixed-pattern noise affecting the red part of the Narval spectra.
This combined spectrum was used for the determination of the apparent atmospheric parameters and for a detailed abundance analysis conducted using the SME software (Spectroscopy Made Easy - version 503) written in the IDL language \citep{1996A&AS..118..595V,2017A&A...597A..16P}.
In our analysis, we did not take into account the non-sphericity and gravity darkening due to rapid rotation, but constrain our investigation to the apparent values.

Following an approach described by \citet{2016MNRAS.456.1221R}, for the fitting of synthetic spectra to the observations six intervals were chosen: 4400--4700, 5100--5250, 5570--5750, 6000--6220, 6400--6700 and 7700--7900 \AA. The 6400--6700 \AA\, region contains the H$\alpha$ line which is a good temperature indicator in the cool stars' domain and is still slightly sensitive to gravity variations at effective temperatures near 7000~K \citep[see][]{2016MNRAS.456.1221R}. 
The 7700--7900 \AA\ region contains the \ion{O}{i} triplet which allows estimation of the oxygen abundance in non-local thermodynamic equilibrium (NLTE). In addition, the small spectral region from 5850--5910\,\AA\, with the \ion{Ba}{ii} line at a wavelength of 5853\,\AA\, and the resonance \ion{Na}{i} D lines was used for abundance determination. 
Atomic line parameters were extracted from the third version of the Vienna Atomic Line Database \citep[VALD3;][]{2015PhyS...90e4005R, 2017ASPC..510..518P}.
Besides O, NLTE effects were taken into account for Na, Ca, Ba \citep{2017ASPC..510..509P}. Strong observed lines of these elements allow measurement of accurate abundances even in such a rapidly rotating star. The SME analysis was performed with the grid of plane-parallel 
%spherically symmetric 
MARCS atmospheric models \citep{2008AA...486..951G}. The library of the departure coefficients for this grid was calculated based on the model atom developed for O by \citet{2013AstL...39..126S}, for Na by \citet{2014AstL...40..406A}, for Ca by \citet{2018MNRAS.477.3343S} and for Ba by \citet{1999A&A...343..519M}. 

\begin{table*}
\begin{center}
\caption{Atmospheric parameters of $\beta$\,Cas used for the determination of the abundance of Nd.}
%\begin{scriptsize}
\begin{tabular}{lrrrrrr}
\hline \hline 
%\multicolumn{7c}{Atmospheric parameters}
 & \multicolumn{3}{c}{This paper} & \multicolumn{2}{c}{\citet{che2011}} & \citet{2003AJ....126.2048G} \\
Parameter            &   Value   &  err$_1$   & err$_2$ & Model 1 & Model 2 &  \\
\hline
\Teff\ [K]    & 6920 & 35  &140 & 6825 & 6897& 6915 \\
\logg\ [cgs]       & 3.53 & 0.16&0.58& 3.57 & 3.59& 3.49 \\
 $[M/H]$    &$-0.11$ & 0.04&0.12&   -   & - &$-0.02$ \\
\vsini\ [\kms] & 73.6 & 8.1 &7.0 & 72.4 & 79.8 &   -   \\
\vmic\ [\kms] & 4.1  & 0.4 &0.5 &   -   &  -   & 3.1  \\
\hline
\end{tabular}
%\end{scriptsize}
\label{param}
\tablefoot{The errors, err$_1$ and err$_2$, were calculated using SME and are based on the confidence interval (method 1) and the statistical analysis of the residuals (method 2). Apparent parameters from modelling by \citet{che2011} are given in columns 5-6. The last column contains parameters derived by \citet{2003AJ....126.2048G}.}
\end{center}
\end{table*}

SME implements two methods for estimating parameter uncertainties when fitting stellar spectra. The first is a standard estimate of the confidence interval based on the covariance matrix. This matrix is just the inverse of the Marquardt-Levenberg Hessian matrix approximation computed at the best fit solution \citep[see, e.g.][Section 15 Modelling of Data]{2002nrca.book.....P}. The covariance matrix is stored as part of the SME output structure. The main diagonal contains the squares of confidence intervals (1$ \sigma$ for normal distribution of uncertainties) for all free parameters of the fit. This is true under the assumption of a perfect model (that is the residuals of the fit gradually go to zero as data accuracy improves). Such a situation will correspond to a reduced $\chi^2$ reaching unity, which is seldom the case in spectral synthesis. In practice, for $\chi^2$ values larger than 1 this method gives highly underestimated values of uncertainties. In our error estimates using method 1, we account for this effect by multiplying the main diagonal numbers by the reduced $\chi^2$. The second more heuristic approach described in \citet{2017A&A...597A..16P} and implemented in \citet{2016MNRAS.456.1221R} is based on the statistical analysis of the residuals. In this case, the focus is on the core of the distribution that, given good statistics, resembles a normal distribution. Therefore, the new method provides very reasonable uncertainty estimates for parameters constrained by essentially all spectral lines (e.g., effective temperature, metallicity, velocities). 
The estimates are less robust in case that few lines are sensitive to the parameter (e.g., surface gravity, individual abundances for species represented by one or a few lines etc.).
%The estimates are less robust in case only a small number of spectral elements is affected (e.g., surface gravity, individual abundances for species represented by one or a few lines etc.). 
For abundance estimates of Sr, Nd, and Eu we applied the spectrum fitting procedure used in the  BinMag6 code \citep{2018ascl.soft05015K}.

Synthetic spectra are calculated with the model atmosphere and abundance table derived by the SME procedure, while Sr, Eu, Nd abundances are varied to reach the best fit. For the Sr and Eu abundance estimates, a fit of the synthetic spectrum to the observed spectrum was performed in the 4200-4230\,\AA\ region which contains the \ion{Sr}{ii} $\lambda$~4215.52~\AA\ and \ion{Eu}{ii} $\lambda$~4205.05~\AA\ lines. Additionally, the \ion{Eu}{ii} line at $\lambda$~6645.11~\AA\ was fit. The spectral region 5200-5350~\AA\ was used for the determination of the abundance of Nd.
This region contains many \ion{Nd}{ii} and the strongest \ion{Nd}{iii} lines, which appear in stars with high overabundance of the rare-earth elements. The error estimates for Sr, Nd, Eu are obtained using method 1.

The final parameters of $\beta$\,Cas derived from the spectral synthesis are \Teff\,=\,6920\,K, \logg\,=3.53 cgs, $[M/H]$\,=\,$-0.11$, \vsini\,=\,73.6\,\kms, and microturbulent velocity, \vmic\,=\,4.1\,\kms. They are listed together with the error estimates from both methods (err$_1$ and err$_2$) and a comparison to literature values in Table \ref{param}. 

As described above, our SME analysis was performed using a grid of plane-parallel 
atmospheric models. The surface integration implemented in SME assumes a spherical star with a homogeneous surface.
However, according to \citet{che2011}, $\beta$\,Cas is a rapidly rotating star spinning close to its critical velocity, and hence has inhomogeneous surface temperature and gravity  distributions. The authors modelled $\beta$\,Cas using two methods, and provided apparent effective temperatures, luminosities and masses. The corresponding effective temperatures and gravities
%estimated from these apparent parameters 
for both models -- the modified von Zeipel model (Model 1) and the Lucy model (Model 2) -- are given in columns 5 and 6 of Table~\ref{param}. Within the errors, our parameters agree with those obtained by \citet{che2011}. They also agree with the parameters derived by \citet{2003AJ....126.2048G}. 

The results of the abundance analysis are presented in Table~\ref{abun}. Columns err$_1$ and err$_2$ list the errors of the abundances according to the two methods for error determination implemented in SME. The last column shows element abundances relative to solar values. The solar abundances are taken from \citet{2015A&A...573A..25S, 2015A&A...573A..26S} and \citet{2015A&A...573A..27G}.

\begin{table}
\caption{Abundances in the atmosphere of $\beta$\,Cas.}
\begin{tabular}{lcccl}
\hline \hline 
Element     & log(N$_{el}$/N$_{tot}$)& err$_1$ &  err$_2$ &[N$_{el}$/N$_{tot}$]  \\
\hline
C*  &~$-$3.52 & 0.10 & 0.26 & $+$0.09   \\
O   &~$-$3.19 & 0.02 & 0.06 & $+$0.16   \\
Na  &~$-$5.87 & 0.03 & 0.12 & $-$0.04   \\
Mg  &~$-$4.57 & 0.06 & 0.16 & $-$0.12   \\ % 0.07(09)
Al* &~$-$5.81 & 0.23 & 0.11 & $-$0.20   \\
Si  &~$-$4.55 & 0.07 & 0.19 & $-$0.02   \\ %-0.07(14)
%S*  &-4.82 & 0.41 & 0.56 & +0.10   \\ %-0.27(06)
Ca  &~$-$5.84 & 0.05 & 0.15 & $-$0.12   \\ %-0.04(26)
Sc  &~$-$8.93 & 0.10 & 0.15 & $-$0.05   \\ % 0.02(28)
Ti  &~$-$7.14 & 0.03 & 0.17 & $-$0.03   \\ %-0.10(08)
Cr  &~$-$6.54 & 0.05 & 0.22 & $-$0.12   \\ %-0.08(10)
Fe  &~$-$4.76 & 0.02 & 0.13 & $-$0.18   \\ % 0.00(10)
Ni  &~$-$6.05 & 0.08 & 0.29 & $-$0.21   \\ %-0.09(11)
%Cu* &-8.27 & 0.44 & 0.27 & -0.41   \\ %-0.27
Sr* &~$-$9.13 & 0.10 &      & $+$0.08   \\   
Y   &~$-$9.75 & 0.09 & 0.18 & $+$0.08   \\ % 0.00(02)
Zr  &~$-$9.40 & 0.19 & 0.37 & $+$0.05   \\
Ba  &~$-$9.74 & 0.03 & 0.06 & $+$0.05   \\ % 0.56(18)
Nd* &$-$10.57 & 0.10 &      & $+$0.05   \\
Eu* &$-$11.40 & 0.10 &      & $+$0.12   \\ 
\hline                                     
\end{tabular}
\label{abun}
\tablefoot{Columns err$_1$ and err$_2$ list the errors of the abundances according to the two methods for error determination implemented in SME. The last column shows the abundances relative to solar values. The solar abundances are taken from \citet{2015A&A...573A..25S,2015A&A...573A..26S} and \citet{2015A&A...573A..27G}. Chemical species marked by asterisks have uncertain abundance values because they have only few weak or blended lines in the analysed parts of the spectrum (see text).
}
\end{table}

Chemical species marked by asterisks in Table \ref{abun} have few lines in the analysed parts of the spectrum; therefore their abundances are very uncertain. Sodium and barium also have few lines but they are strong enough to provide a reasonable abundance estimate. Abundances of heavy elements are rather uncertain, too, but are close to the solar values within their errors. The atmosphere of 
$\beta$\,Cas is slightly metal deficient in iron peak elements and slightly overabundant in both light elements C and O, and heavy elements. Overall, the observed abundance pattern of $\beta$\,Cas is similar to the atmospheric abundances of another $\delta$~Scuti-type star HD~261711 \citep{2013A&A...552A..68Z}.

\begin{figure}
\begin{center}
\includegraphics[width=0.48\textwidth]{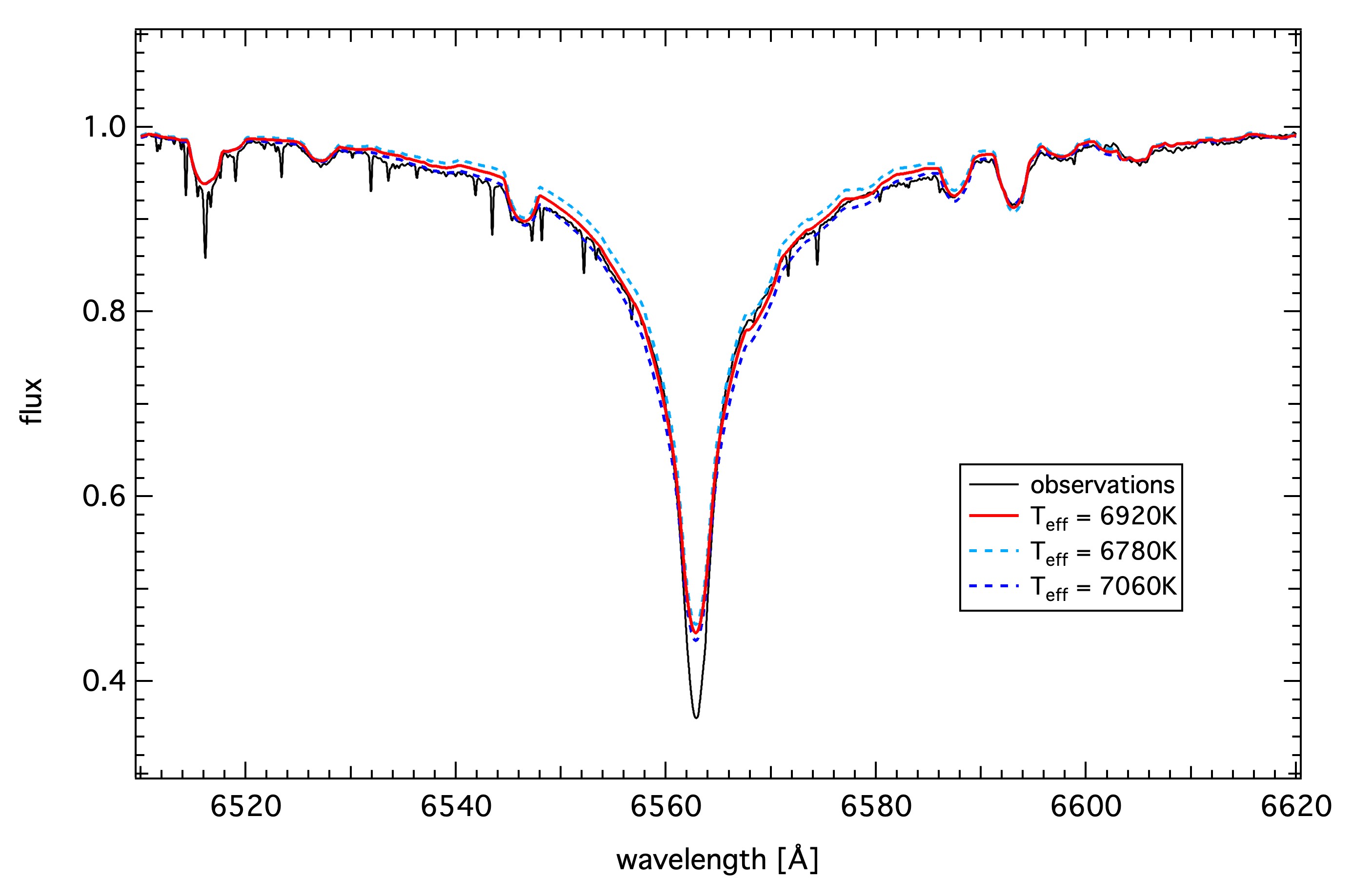}
\caption{Region of the H$\alpha$ line for $\beta$\,Cas: the observed spectrum is shown in black and the calculated synthetic spectrum with the final adopted parameters of \Teff\,=\,6920\,K and \logg\,=\,3.53\,cgs in red. The dashed light blue line represents a synthetic spectrum with a 140\,K lower \Teff, and the dashed dark blue line a synthetic spectrum with a 140\,K higher \Teff.}
\label{fig:Halpha}
\end{center}
\end{figure}

\begin{figure}
\begin{center}
\includegraphics[width=0.48\textwidth]{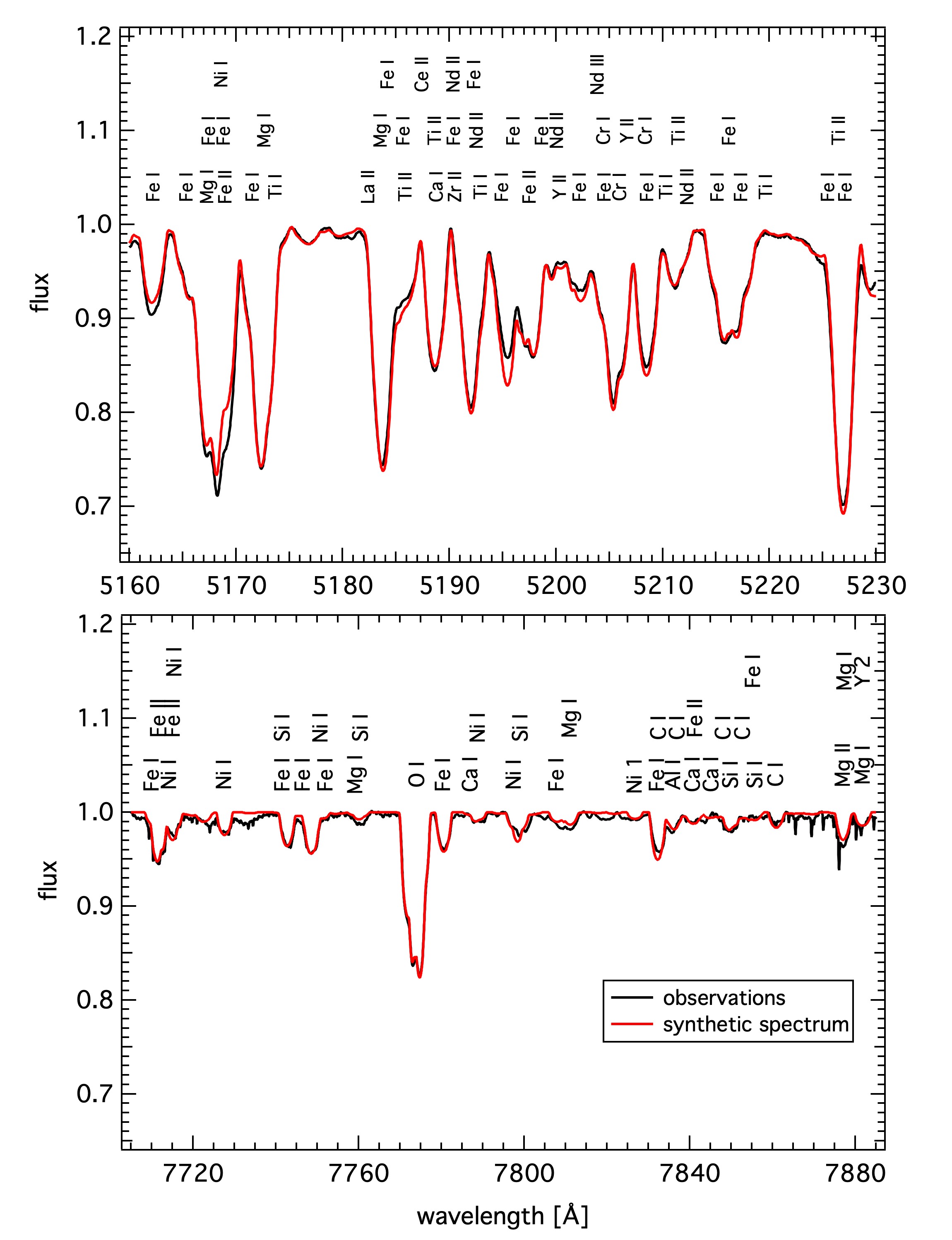}
\caption{Regions from 5160 to 5240\,\ang\, (top panel) and 7705 to 7885\,\ang\ (bottom panel): the observed spectrum is shown in black and the calculated synthetic spectrum with the final adopted parameters of \Teff\,=\,6920\,K and \logg\,=\,3.53\,cgs in red. }
\label{fig:spec2}
\end{center}
\end{figure}

Figures \ref{fig:Halpha} and \ref{fig:spec2} illustrate our best fitting solution using the derived fundamental parameters and atmospheric abundances in the region around H$\alpha$ and in two selected regions.

%{\color{red} If I understand correctly, you did not take into account the non-sphericity and gravity darkening due to rapid rotation. Maybe we should add a few words to explain how these effects could influence the fundamental parameters.}

\section{Spectropolarimetric analysis}

\subsection{LSD profiles}
\label{sect:LSD}
%Part by Coralie

We used the Least Squares Deconvolution (LSD) technique \citep{donati1997} to create mean Stokes I, Stokes V, and N profiles of each Narval polarimetric sequence. The velocity step used for the LSD profiles is 2.6 km~s$^{-1}$.

To perform LSD, a list of stellar lines present in the spectrum, together with their wavelength, depth, and Land\'e factor, is necessary. We started from a line list extracted from the VALD3 atomic database for the effective temperature, gravity, microturbulence, and chemical abundances of $\beta$\,Cas determined above (Table \ref{param}) and restricted it to lines with depths larger than 1\% of the continuum level. From this template list we rejected the hydrogen lines, as well as lines blended with hydrogen line wings, lines that were not present in the spectrum, or that were contaminated by telluric or interstellar features. We then adjusted the line depths given in the line list to those observed in the real spectrum. This method is described in more detail by \cite{grunhut2017}. The final line list includes 5104 lines, with a mean wavelength of 526.47 nm and a mean Land\'e factor of 1.2. 

LSD profiles obtained on the same night were averaged. When 10 or 15 spectra were available for the same night, averages of the LSD profiles by group of 5 consecutive observations (as presented in Table~\ref{logNarval}) have been computed. We finally obtained 19 LSD profiles with typical effective integration time of about 40~min each. These averaged LSD profiles have a S/N ranging from 68000 to 125000 in Stokes V and N, and 5000 to 10000 in Stokes I. 
Representative average LSD profiles are shown in Fig.~\ref{fig:selected-lsd}. A complete set of Stokes I, V, and N LSD profiles is presented in appendix in  Fig.~\ref{all-lsd}.

\begin{figure}
\begin{center}
\includegraphics[width=\hsize]{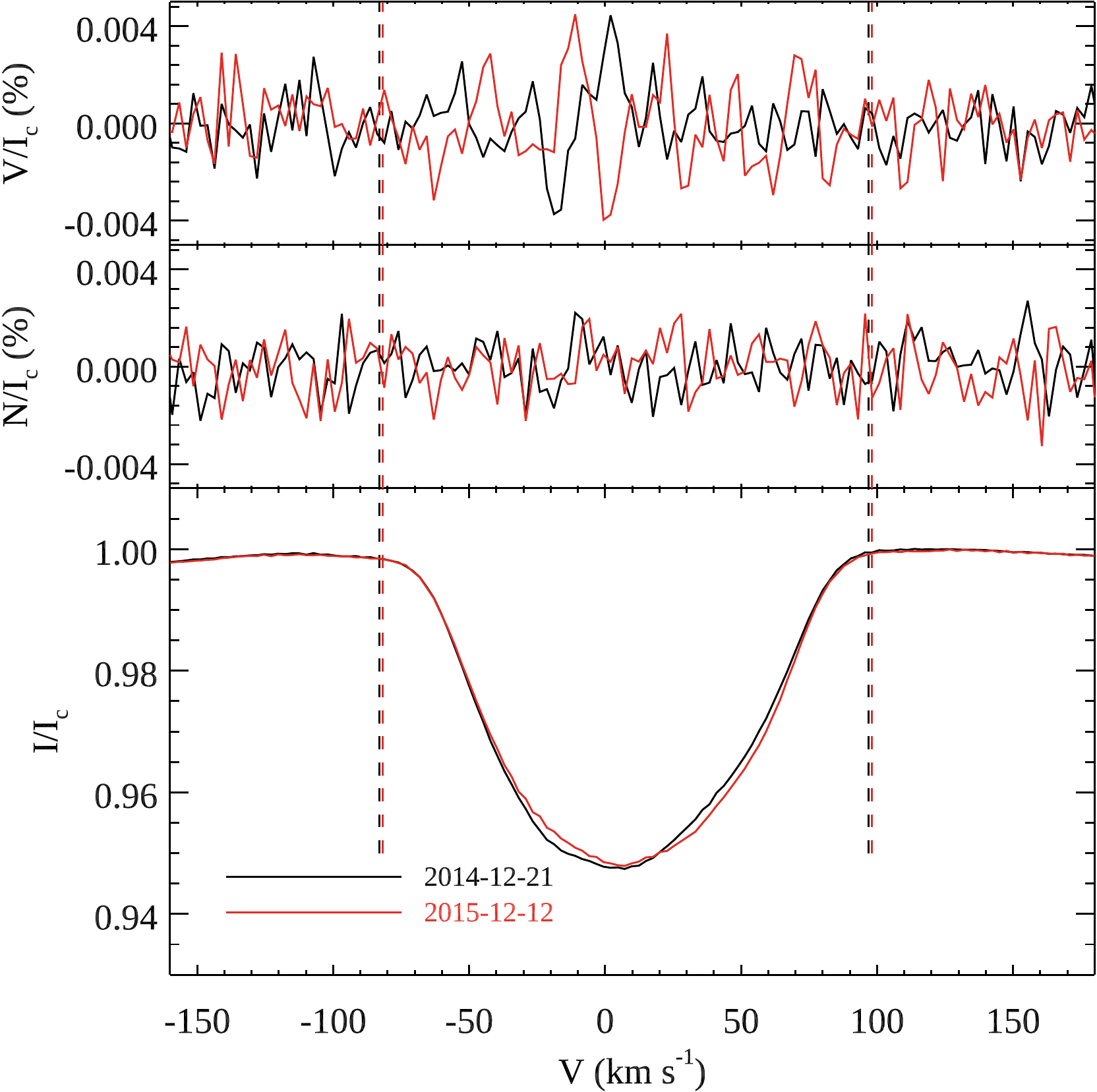}
\caption[]{Representative LSD Stokes I, V, and null profiles of $\beta$\,Cas. The two sets of profiles shown here correspond to the Narval observations obtained on 2014-12-21 (data set \#4 in Table~\ref{logNarval}) and 2015-12-12 (data set \#18). The vertical dashed lines indicate measurement windows adopted in Sect.~\ref{sect:5.1} for these two observations.}
\label{fig:selected-lsd}
\end{center}
\end{figure}

The LSD Stokes I profiles show clear variations related to the $\delta$\,Scuti pulsations. The LSD Stokes V profiles show weak and rather complex Zeeman signatures, while N does not show such a signal. This is a clear indication of the presence of a magnetic field in $\beta$\,Cas. 

\subsection{Magnetic analysis}
\label{sect:5.1}
% Part by Coralie

\begin{table}
\caption{Magnetic field measurements for $\beta$\,Cas. }
\begin{tabular}{lrrc}
\hline \hline 
\# & $B_l$~~~~~~ & $N_l$~~~~~~ & Magnetic detection \\
\hline
1& 0.6	  $\pm$6.6	&     $-$1.1 	  $\pm$6.6 & ND \\
2& 1.3	  $\pm$2.5	&     $-$4.6 	  $\pm$2.5 & MD \\
3& 3.9	  $\pm$1.9	&     $-$0.0  	  $\pm$2.0 & DD \\
4& 0.8	  $\pm$2.1	&     1.4  	  $\pm$2.1 & DD \\
5& $-$5.9   $\pm$2.5	&     $-$2.3 	  $\pm$2.5 & DD \\
6& 0.6	  $\pm$2.7	&     1.6  	  $\pm$2.7 & DD \\
7& $-$1.6   $\pm$3.4	&     2.1  	  $\pm$3.4 & ND \\
8& $-$1.4   $\pm$3.0	&     $-$1.7 	  $\pm$3.0 & MD \\
9& $-$3.5   $\pm$2.6	&     $-$1.3 	  $\pm$2.6 & MD \\
10& $-$3.1  $\pm$2.2	&     $-$2.6 	  $\pm$2.2 & DD \\
11& 1.0	  $\pm$2.1	&     0.6  	  $\pm$2.1 & MD \\
12& $-$0.1  $\pm$2.1	&     0.0  	  $\pm$2.1 & DD \\
13& $-$3.2  $\pm$2.1	&     1.0  	  $\pm$2.1 & ND \\
14& $-$3.7  $\pm$2.3	&     0.7  	  $\pm$2.3 & DD \\
15& 3.4	  $\pm$2.6	&     $-$3.9 	  $\pm$2.6 & DD \\
16& 2.8	  $\pm$2.2	&     $-$1.8 	  $\pm$2.3 & DD \\
17& $-$0.3  $\pm$2.4	&     $-$1.4 	  $\pm$2.4 & DD \\
18& $-$0.9  $\pm$2.7	&     $-$0.6 	  $\pm$2.7 & DD \\
19& 1.7	  $\pm$2.4	&     $-$2.7 	  $\pm$2.4 & ND \\
\hline
\end{tabular}
\label{Bl}
\tablefoot{Columns indicate the number of the polarimetric measurement, the longitudinal field value with its error bar in Gauss, the corresponding measurement obtained for N, and the formal magnetic field detection status (DD is a Definite Detection, MD is a Marginal Detection, and ND is No Detection) based on Stokes V profile analysis.}
\end{table}

%Since the $B_l$ values are all compatible with 0, the 
The detection of a magnetic field in each measurement was formally evaluated by the False Alarm Probability (FAP) of a signature in the Stokes V profile inside the velocity domain defined by the Stokes I line width (i.e. $\pm90$  km~s$^{-1}$ around the line centre), compared to the mean noise level in the Stokes V profile outside this domain. We adopted the convention defined by \cite{donati1997}: if FAP $<$ 0.001\%, the magnetic detection is definite (DD), if 0.001\% $<$ FAP $<$ 0.1\% the detection is marginal (MD); otherwise there is formally no magnetic detection (ND). The detection status for each average profile is reported in Table~\ref{Bl}. 

We obtain 11 DD, 4 MD, and 4 ND in Stokes V inside the spectral line, while there is no detection outside the line and no detection in the N profiles. This confirms that a magnetic field is repeatedly detected at the surface of  $\beta$\,Cas. 

% OK: moved this paragraph from the beginning of this section to here
Using the LSD Stokes V and I profiles, and a center-of-gravity method \citep{rees1979,wade2000}, we calculated the longitudinal magnetic field value $B_l$ corresponding to the observed Zeeman signatures in the velocity range defined by the Stokes I line width, i.e. $\pm90$  km~s$^{-1}$ around the line center. We applied the same calculation to the N profiles to obtain $N_l$. These values are reported in Table~\ref{Bl} for the averaged profiles. We find that the longitudinal field values are very weak, of the order of a few Gauss. 
% OK: added this sentence
All $B_l$ that we derive are compatible with 0, indicating a complex magnetic field topology typical of cool active stars with dynamo fields \cite[e.g.][]{donati1997,donati_abdor}.

\subsection{Zeeman Doppler Imaging}
\label{sect:ZDI}

We used Zeeman Doppler imaging \citep[ZDI,][]{kochukhov2016} to infer the magnetic field topology of $\beta$\,Cas and to put constraints on the stellar rotational period. The tomographic mapping was carried out with the help of the magnetic inversion code {\sc InversLSD} \citep{kochukhov2014}, which was modified to include a non-radial pulsation velocity field using spherical harmonic parameterisation described by \citet{kochukhov2004}. In addition, we lifted the usual approximation of ZDI that each observation corresponds to an instantaneous profile measurement. Instead, the model line profiles were calculated by a numerical integration over appropriate pulsational and rotational phase intervals, which allowed us to properly account for phase smearing. On the other hand, the gravity darkening and non-sphericity resulting from a rapid rotation of $\beta$\,Cas were not included.

A subset of observations obtained in the period from 01 Dec 2015 to 13 Dec 2015 was considered for the ZDI modelling. This data set comprises 75 individual Stokes $V$ observing sequences ($4\times65$~s each), in groups of 5 consecutive observations repeated 1--3 times per night. In the first analysis step we modelled all 75 Stokes $I$ LSD profiles in order to optimise \vsini\ together with the pulsational mode parameters. This was accomplished with the help of a series of forward calculations assuming a single-mode pulsation with $P_{\rm puls}=0.1010396$~d (i.e., the average period corresponding to our F$_1$) and $i=20\degr$ \citep{che2011}. The pulsational velocity amplitude and phase were adjusted to reproduce the observed RV curve. Based on this analysis, we inferred \vsini\,=\,75~km\,s$^{-1}$, which is close to the value of 73.6~km\,s$^{-1}$ found in Sect.~\ref{Sec:spectro}, and found that an axisymmetric quadrupolar mode provides a marginally better description of the pulsational Stokes $I$ profile variability pattern compared to the radial or axisymmetric dipolar pulsation. 

In a second step we used the previously determined broadening and pulsational parameters to reconstruct the magnetic field geometry of $\beta$\,Cas for different trial values of the stellar rotational period. This modelling was based on the 15 Stokes $V$ averaged spectra obtained in 2015 (i.e. numbers \#5 to \#19 in Table~\ref{logNarval}).

\begin{figure}[!t]
\centering
\resizebox{\hsize}{!}{\rotatebox{0}{\includegraphics{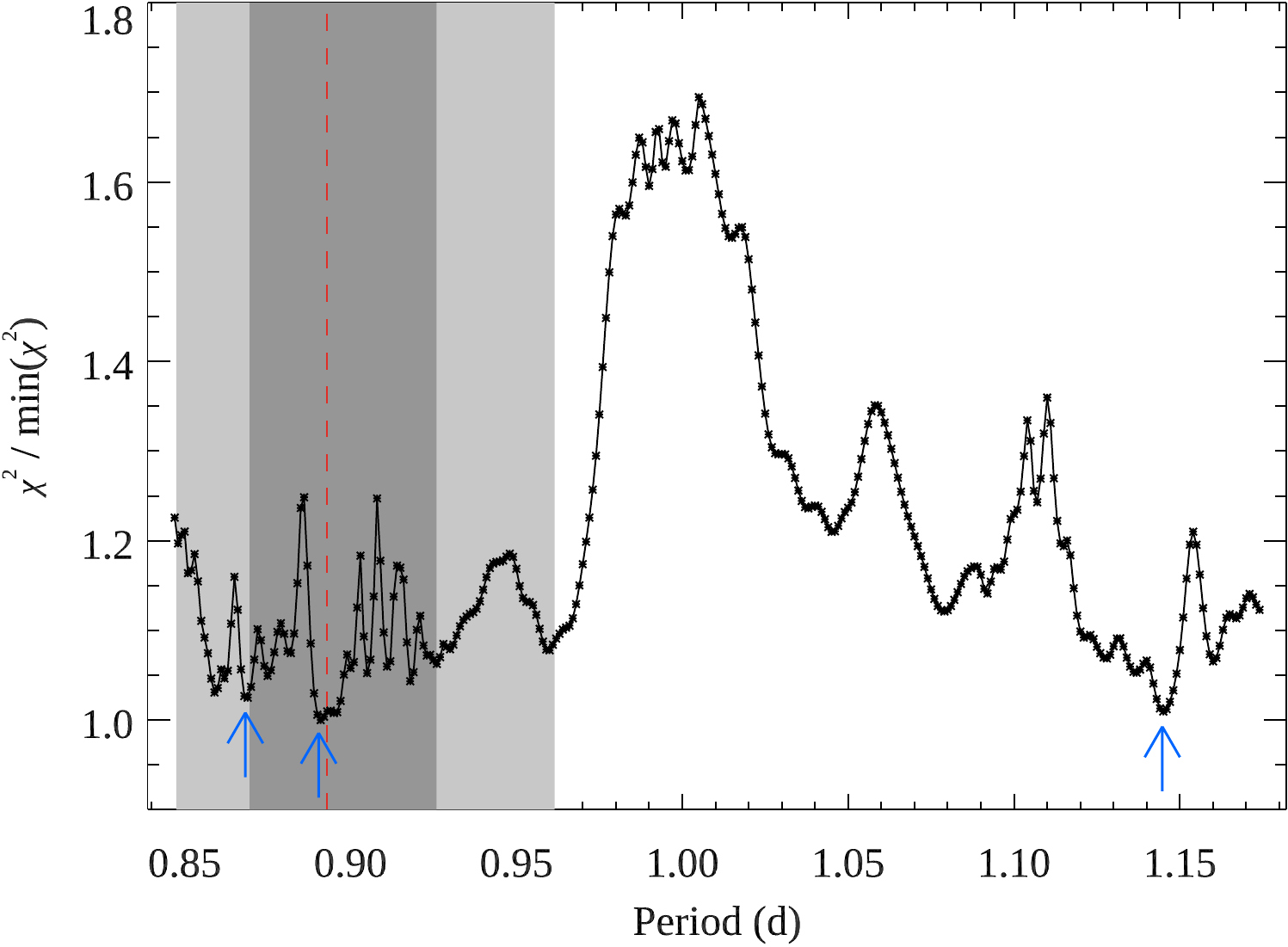}}}
\caption{Relative $\chi^2$ of the fit to Stokes $V$ LSD profiles as a function of rotational period. The vertical dashed line and the shaded regions correspond to the rotational period determined by \citet{che2011} and the corresponding 1--2$\sigma$ error bars. The arrows indicate the three rotational periods discussed in the text.}
\label{fig:prot}
\end{figure}

\begin{figure}[!th]
\centering
\resizebox{\hsize}{!}{\rotatebox{0}{\includegraphics{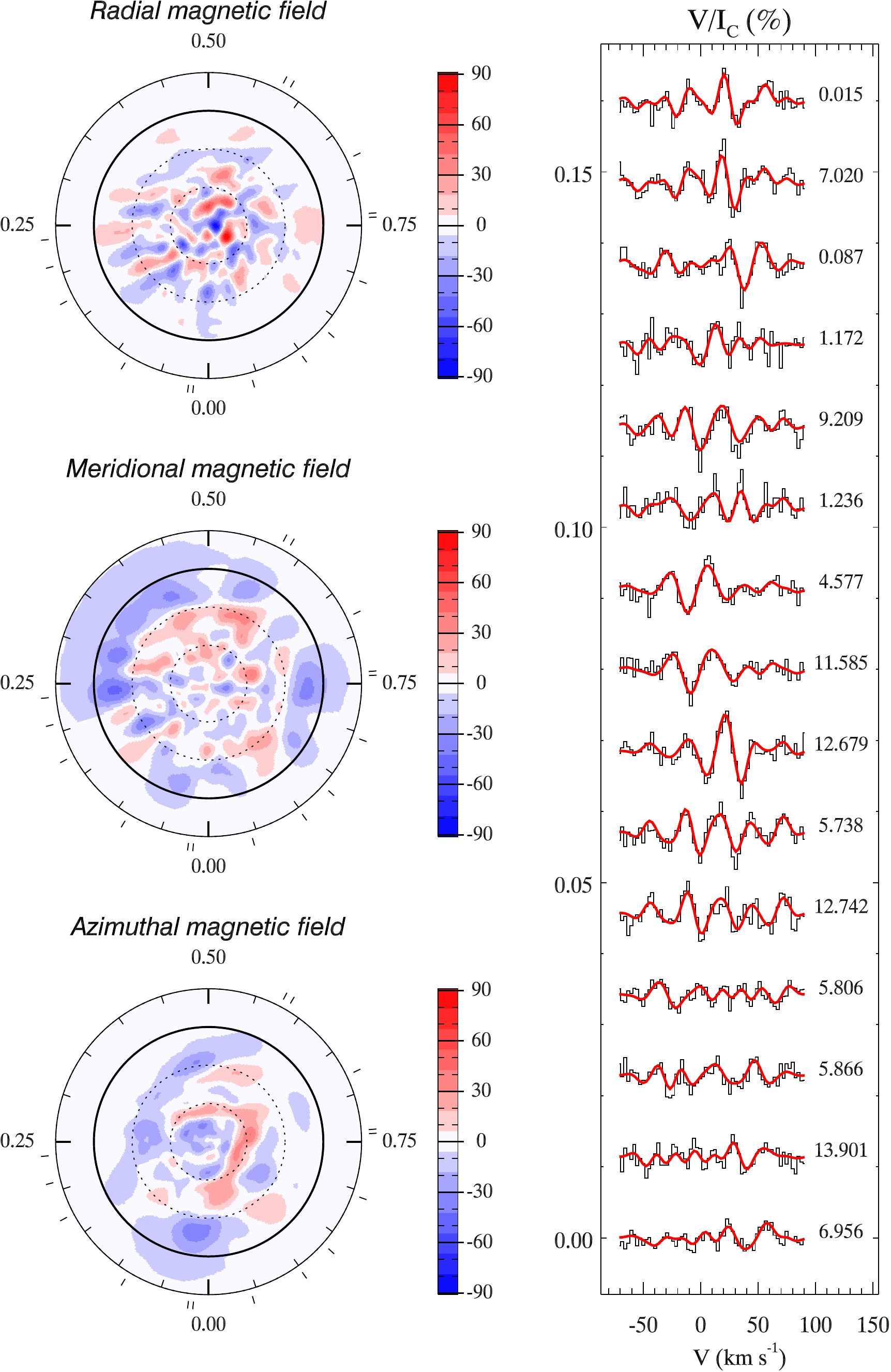}}}
\caption{Magnetic field maps of $\beta$\,Cas and corresponding Stokes $V$ profile fits obtained with ZDI for $P_{\rm rot}=0.868$~d. The plots on the left side show the radial, meridional, and azimuthal magnetic field components in the flattened polar projection. The thick circle corresponds to the stellar equator. The numbers next to surface plots correspond to rotational phases while the short bars illustrate rotational phase coverage. The colour bars indicate the field strength in Gauss. 
The right panel shows the observed (histogram) and model (solid red curves) profiles, shifted vertically with an equidistant step. The rotational phases (calculated relative to HJD$_0$\,=\,2457358.31370) are indicated to the right of each line profile.}
\label{fig:zdi}
\end{figure}

\begin{figure}[!th]
\centering
\resizebox{\hsize}{!}{\rotatebox{0}{\includegraphics{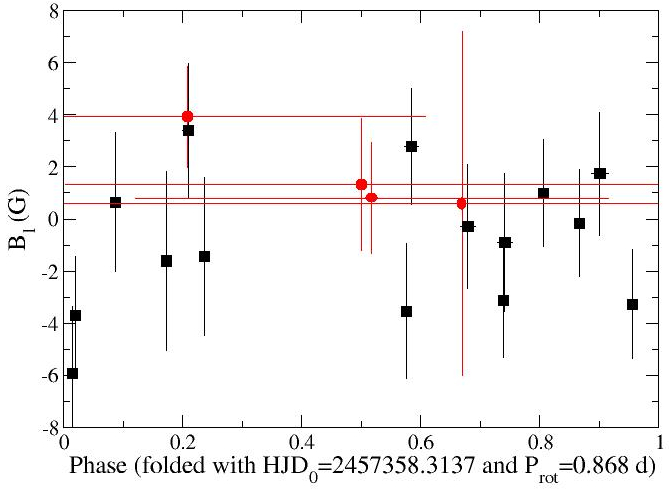}}}
\caption{Longitudinal magnetic field values folded with the rotation period 0.868 d, with phases calculated relative to  HJD$_0$\,=\,2457358.3137 corresponding to the start of the 2015 observations. The red symbols indicate the data obtained in 2013 and 2014, which thus have a much larger phase error compared to the 2015 data set.}
\label{fig:bl}
\end{figure}

\citet{che2011} estimated the rotational frequency to be $1.12\substack{+0.03 \\ -0.04}$ d$^{-1}$. This corresponds to $P_{\rm rot}=0.89\substack{+0.03 \\ -0.02}$~d. For the ZDI modelling, we considered a $P_{\rm rot}$ interval of 0.847--1.174~d, which encompasses the $\pm2\sigma$ rotational period range from \citet{che2011} and extends all the way to 1.172~d corresponding to the difference between the two main frequencies present in the BRITE data. The resulting relative $\chi^2$ of the fit to Stokes $V$ profiles is illustrated as a function of trial rotation period in Fig.~\ref{fig:prot}. We found that multiple rotation periods provide good descriptions of our Stokes $V$ observations of $\beta$\,Cas. Specifically, the lowest $\chi^2$ of the fit to the observed LSD profiles is achieved with $P_{\rm rot}$ of 0.868, 0.890, and 1.145~d. All three rotation periods result in qualitatively similar magnetic field maps and a non-sinusoidal behaviour of the $B_l$ values. The ZDI field geometry and Stokes $V$ profile fits corresponding to $P_{\rm rot}=0.868$~d are illustrated in Fig.~\ref{fig:zdi}. The $B_l$ values folded with this period are shown in Fig.~\ref{fig:bl}.

The three rotation periods mentioned above result in a qualitatively different phase distribution of the 15 Stokes $V$ observations. The period $P_{\rm rot}=0.868$~d yields three pairs of Stokes $V$ profiles, obtained 7 nights apart, with very similar rotational phases. These profiles closely agree with each other both in the observations and in the model (see Fig.~\ref{fig:zdi}). This is unlikely to be a coincidence. On the other hand, periods $P_{\rm rot}=0.890$~d and $P_{\rm rot}=1.145$~d yield no pairs of close rotational phases. These periods might therefore represent aliases appearing due to a relatively sparse rotational phase sampling of our observations. For this reason, we consider $P_{\rm rot}=0.868$~d to be the most likely rotational period of $\beta$\,Cas. The ZDI inversion with this period also provides a magnetic map with 14\% lower total energy compared to the reconstruction results for the other two periods. Formally, $P_{\rm rot}=0.890$~d  cannot be excluded as it produces a low $\chi^2$ and is consistent with the results of \citet{che2011}. However, it does not explain the observed repetition of Stokes $V$ profile shapes. The period $P_{\rm rot}=1.145$~d has the same problem; in addition it is formally excluded (at the 8.4$\sigma$ level) by the interferometric results of \cite{che2011}.

For each of the three plausible periods we tested the possibility of the presence of a solar-like differential rotation. In all three cases equally good fits can be obtained with no differential rotation ($\alpha\equiv \Delta\Omega/\Omega_{\rm e}=0.0$) and with a combination of a shorter equatorial period and a moderate solar-like differential rotation ($\alpha$\,=\,0.01--0.02). In other words, the existing observational data do not allow us to meaningfully constrain differential rotation.

Considering the ZDI results obtained for $P_{\rm rot}=0.868$~d, we performed a bootstrapping error analysis by randomly reshuffling residuals between observations and model fit for each rotational phase, adding these residuals back to the model profiles and reconstructing a new magnetic map from simulated data. This procedure was repeated 100 times and the resulting standard deviations of different magnetic parameters were adopted as error estimates. This analysis showed that the local error of the magnetic field vector maps shown in Fig.~\ref{fig:zdi} is 6~G for the radial field and 8--9~G for the meridional and azimuthal components. The maximum surface field strength is $87\pm5$~G while the mean field strength (field modulus averaged over the visible hemisphere) is $20\pm1$~G. 

The spherical harmonic description of the surface magnetic field  implemented in {\sc InversLSD} (see \citealt{kochukhov2014} for details) allows us to assess contributions of different modes to the field geometry of $\beta$\,Cas. We infer that the magnetic field of this star is predominantly poloidal ($65\pm5$\% of the field energy is concentrated in poloidal modes) and contains a comparable contribution of axisymmetric ($|m|<\ell/2$, $60\pm5$\%) and non-axisymmetric ($|m|\ge \ell/2$, $40\pm5$\%) harmonic components. 
Distribution of the relative magnetic field energy as a function of $\ell$ value of harmonic modes is shown in Fig.~\ref{zdi-ell}. We find that the field energy peaks at $\ell=1$ ($14\pm6$\% of the total energy) and then fluctuates between 3--4\% until $\ell=10$. Then, there is a secondary maximum at $\ell=12$--15 (energy contributions up to $5.4\pm0.7$\% per $\ell$ value), corresponding to the small-scale structure seen in the reconstructed magnetic maps. The energy contribution of all modes with $\ell\le27$ appears to be non-negligible ($\ga$\,$1\pm0.2$\% of the total magnetic field energy). Given the high \vsini\ of the star and the resolving power of our spectra, ZDI is potentially sensitive to modes with $\ell$ up to $\approx100$ \citep{fares2012}.

\section{Discussion}

\subsection{Pulsations}

$\delta$ Scuti pulsators are intermediate mass stars located in the lower part of the so-called classical instability strip with spectral types between A2 and F2 \citep{rodriguez2001} that can be in the pre-main sequence, main sequence, or post-main sequence evolutionary stages. $\delta$ Scuti stars are typically multi-periodic oscillators that can show very rich pulsation frequency spectra
%\citep[e.g.,][]{poretti2009} 
that challenge asteroseismic analyses and the theoretical interpretation of the observed pulsation frequencies. 

A and F type stars can also show a combination of pressure ({\it p}-), gravity ({\it g}-), and Rossby ({\it r}-) modes \citep[e.g.,][]{uytterhoeven2011,saio2018}. Hence, we also investigated all available photometric data sets {of $\beta$\,Cas} for the presence of {\it g}- and {\it r-} modes in addition to the already described {\it p}-modes. {\it g}-modes would be found in the range of $\sim$\,3.3 to 0.3\,\cd \ in frequency \citep[corresponding to $\sim$0.3 to 3.0\,days in period;][]{kaye1999}; {\it r-}modes appear as a group of peaks below {\it m} times the rotational frequency. BRITE-Constellation and SMEI data are known to allow the detection of {\it g}-mode pulsations when the standard procedures for data reduction are applied \citep[e.g.,][]{kallinger2017,zwintz2017}. The BRITE-Constellation and SMEI data for $\beta$\,Cas do not show any signs for {\it g}-mode pulsations. We also checked the TESS data for the potential presence of {\it g}-modes before we applied the Gaussian filter to remove the instrumental signal, and also did not find any evidence for them (see Fig.~\ref{fig:TESS-zoom}).
In all our data sets, we do not detect a frequency that can be attributed to the rotation of $\beta$\,Cas or a group of peaks typical for {\it r}-modes. 
%Hence, we also do not find any signs for r-modes in our data for $\beta$ Cas.

%, but did not detect any. 

%As $\beta$\,Cas is an F-type star, it might also be expected to show $g$-mode pulsations. But our investigation did not detect any pulsational signal in the $g$-mode regime, {\bf i.e., from $\sim$0.3 to 3.0\,days in period \citep[or from $\sim$\,3.3 to 0.3\,\cd \ in frequency; ][]{kaye1999}. }

Two of the most prominent physical effects that complicate the asteroseismic analysis of $\delta$ Scuti stars, are (i) moderate to fast rotation which can cause a splitting of the pulsation modes with the same $n$ and $\ell$ values but different $m$ values \citep[e.g.,][]{lignieres2006}, and (ii) the presence of magnetic fields which is a rather recent discovery for $\delta$ Scuti stars \citep[e.g.,][]{neiner-lampens2015,neiner2017}.

$\beta$\,Cas is a special case in this context because observationally it shows only three independent {\it p}-mode pulsation frequencies and not several dozen to hundreds of modes. The reason might be connected to the inclination angle of only 19$^{\circ}$ \citep{che2011} and the presence of the magnetic field. 
Despite $\beta$\,Cas' rather high rotational speed, no sign of a rotational splitting can be detected in the present observational material. 

We calculated the pulsation constant, $Q$, based on our values for the three independent frequencies, F$_1$, F$_2$, and F$_4$, and our fundamental parameters, \Teff\,=\,6920\,K and \logg\,=\,3.53\,cgs. The pulsation constant $Q$ \citep{petersen1976,stellingwerf1979} can be expressed as:
\begin{equation}
    {\rm log}\,Q = -6.454 + {\rm log}\,P + 0.5\,\logg +0.1\,M_{\rm bol} + {\rm log}\Teff ,
\end{equation}
where $P$ is the pulsation period in days, and $M_{\rm bol}$ is the bolometric magnitude, which is defined as $M_{\rm bol} = M_V + {\rm BC}$. The absolute magnitude of 1.14\,mag is calculated from the distance of 16.8\,pc, and the bolometric correction, BC, can be calculated from the relation given by \citet{reed1998}.

\begin{equation}
\begin{split}
BC = -8.499\,({\rm log}\Teff - 4)^4 + 13.421\,({\rm log}\Teff -4)^3 - \\
8.131\,({\rm log}\Teff - 4)^2 -3.901\,({\rm log}\Teff-4) - 0.438.
\end{split}
\end{equation}

The pulsation constant has a value of 0.033 for fundamental mode pulsation ($n = 0$), 0.025 for the first overtone ($n = 1$), 0.020 for the second overtone ($n = 2$), and 0.017 for the third overtone ($n = 3$).

The corresponding $Q$ values for F$_1$ at 9.89708\cd, F$_2$ at 9.0437\cd, and F$_4$ at 8.3847\cd\ are 0.018, 0.020, and 0.022, respectively. The calculated $Q$ value for F$_4$ lies between the first and second overtone pulsation mode; no clear identification of the radial order is therefore possible. 
F$_2$ can be identified as second overtone, and F$_1$ possibly as third overtone pulsation. 
The latter is in contradiction with earlier reports in the literature that identified F$_1$ as a first overtone pulsation mode \citep{rodriguez1992}. Together with the results from our Stokes $I$ LSD profile analysis (see Sect. \ref{sect:ZDI}), F$_1$ would then be a possible $n=3$, $\ell = 2$, and $m=0$ p-mode.

%{\color{red} How does this match the ZDI result that F1 corresponds to l=2, m=0?}

\subsection{Amplitude variability}
\label{sect:ampvar}
%Include: Does the amplitude of the two main pulsation modes change with time? Describe the different passbands of BRITE B, R, and TESS and interpret the different amplitudes qualitatively.

We have used data for $\beta$\,Cas obtained in four specific passbands: the BRITE blue filter (390 -- 460\,nm), the BRITE red filter (550 -- 750\,nm), the passband of SMEI (450 -- 950\,nm), and the passband of TESS (600 -- 1000\,nm). For each of the two pulsation frequencies that appear in data from all instruments, it is evident that the by far highest amplitude can be measured from the BRITE blue filter data, followed by the BRITE red amplitude and the amplitude in the SMEI passband. The pulsation amplitudes are smallest in the reddest passband -- that of the TESS camera. 

It has been reported several times in the past that some of the pulsation amplitudes in $\delta$ Scuti stars show time-dependent behaviour \citep[e.g.,][]{breger1991,breger2002,zwintz2019}. The origin of amplitude variability can be either intrinsic due to beating of unresolved frequencies, non-linearity, or mode-coupling, or extrinsic due to binary and multiple systems. A detailed overview of the amplitude modulation in $\delta$ Scuti stars is given by \citet{bowman2016}. 

\begin{figure}[htb]
\begin{center}
\includegraphics[width=0.48\textwidth]{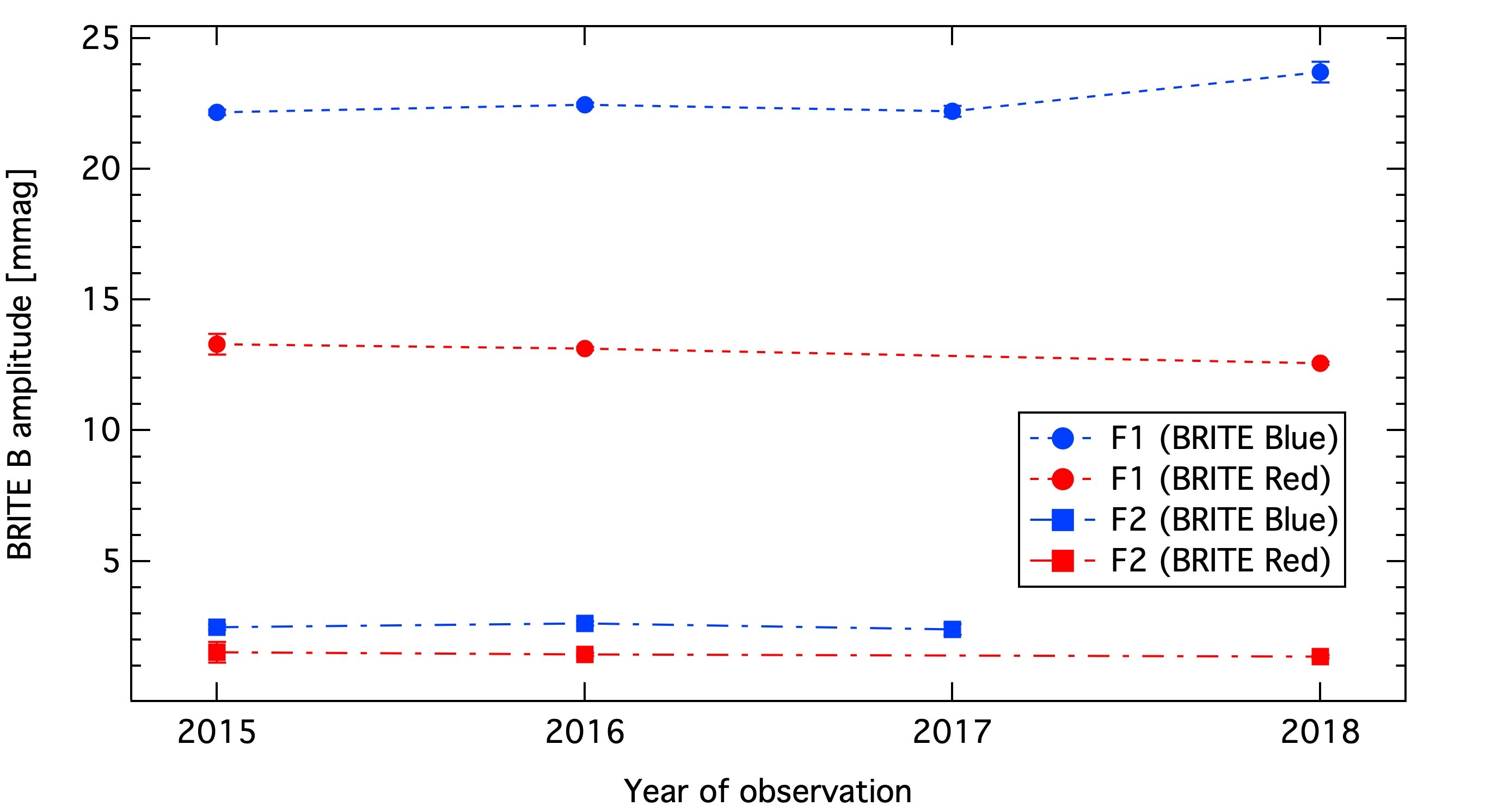}
\caption[]{Annual behaviour of the amplitudes of the two main pulsation frequencies, F$_1$ (filled circles) and F$_2$ (filled squares) based on the BRITE-Constellation observations in the $B$ (blue) and $R$ (red) filter conducted every year from 2015 to early 2019. The one-sigma error bars in the figure are mostly smaller than the symbol size.}
\label{fig:AnnualAmpvar}
\end{center}
\end{figure}

We examined the amplitude variability of $\beta$\,Cas by first studying the annual behaviour in the BRITE red and blue filters based on the four years of consecutive observations (see Table \ref{tab:obs}). It can clearly be seen from Fig.~\ref{fig:AnnualAmpvar} that there is no significant modulation of the amplitudes of F$_1$ and F$_2$ in both BRITE filters.

\begin{figure}[htb]
\begin{center}
\includegraphics[width=0.48\textwidth]{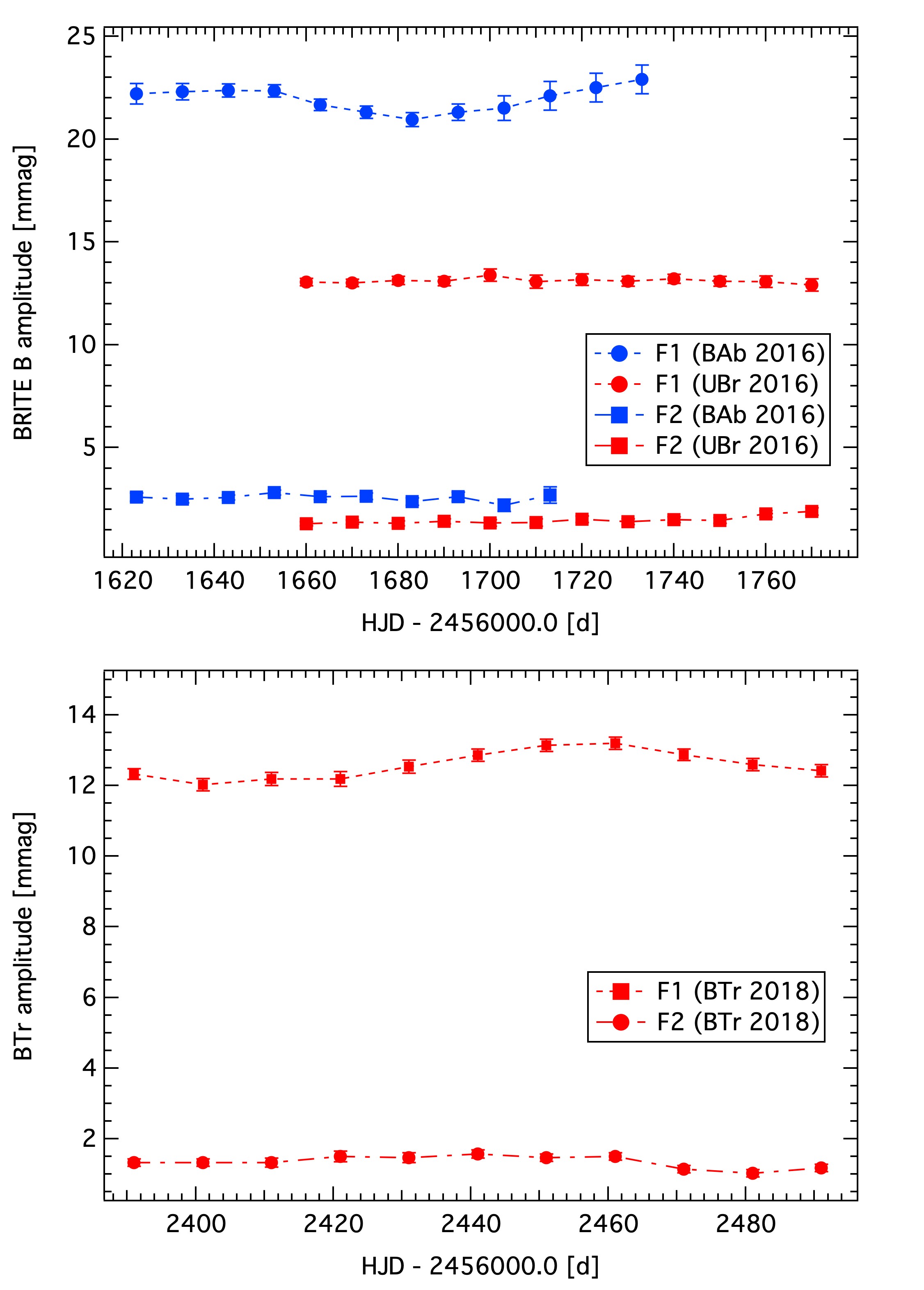}
\caption[]{Seasonal amplitude behaviour of the two main pulsation frequencies, F$_1$ (filled circles) and F$_2$ (filled squares). Top panel: 30-day subsets with 20-day overlaps for the BAb 2016 (blue) and UBr 2016 (red) data  (top panel); bottom panel: 30-day subsets with 20-day overlaps for the BTr 2018 data (red). The one-sigma error bars in the figure are often smaller than the symbol size.}
\label{fig:SeasonalAmpvar}
\end{center}
\end{figure}

In a second step we used the longest and best BRITE-Constellation data obtained in single observing seasons to study the behaviour of the pulsation amplitudes on the shorter time scale of a few months. The first season of BRITE observations of $\beta$\,Cas in 2015 yielded too short time bases (i.e., $\sim$58 days) for such an analysis. The BAb data obtained in 2017 and in 2018 have insufficient quality for such an analysis. Therefore, we restrained the analysis of shorter period amplitude variability on the BAb und UBr data obtained in 2016 and the BTr data from 2018. 
For each of the three data sets, we calculated 30-day subsets with 20-day overlaps, and find only very moderate modulations of the pulsation amplitudes of F$_1$ and F$_2$ within one-sigma errors (see Fig.~\ref{fig:SeasonalAmpvar}). 

%Frequency F$_1$ seems to show some periodicity in the BAb 2016 and the BTr 2018 data. We therefore tried to see if a common period can be found that explains the moderate amplitude variability seen in Fig.~\ref{fig:SeasonalAmpvar}. The amplitudes of F$_1$ in the BAb 2016 data vary with a period of $\sim$91 days, the variability in the BTr 2018 data yields a period of $\sim$101 days. As the BAb 2016 data set has a time base of 145 days and the BTr 2018 data set a time base of 129 days, the data only cover slightly more than one period of any potential amplitude variation. For completeness, we also calculated the period of a potential amplitude variability of F$_1$ using the UBr 2016 data, but no significant period was determined. Therefore, based on the available observational material, we cannot find an unambiguous value for a potential modulation of the pulsation amplitudes for F$_1$.

\begin{figure}[htb]
\begin{center}
\includegraphics[width=0.48\textwidth]{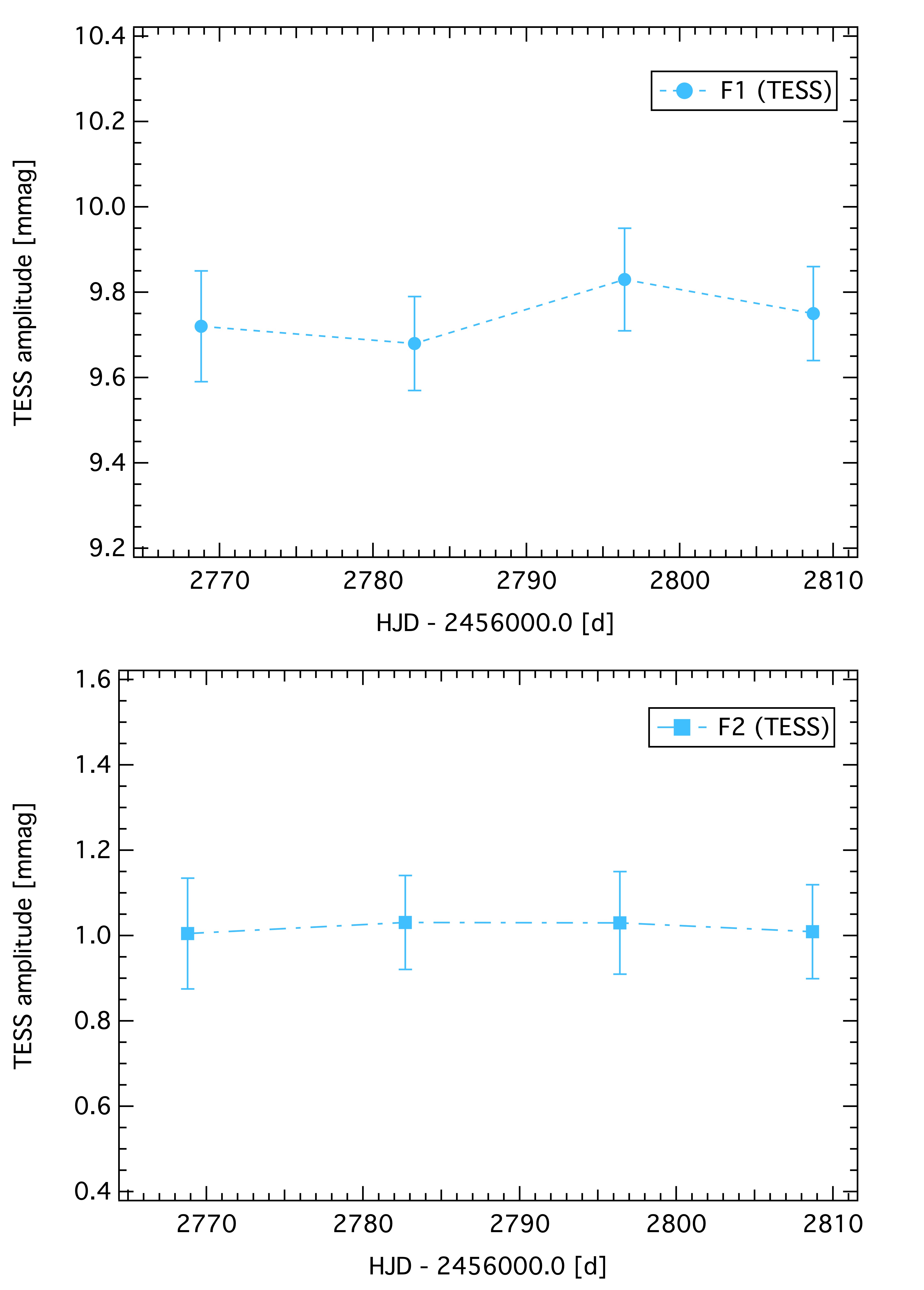}
\caption[]{Amplitude behaviour of the two main pulsation frequencies, F$_1$ (filled circles, top panel) and F$_2$ (filled squares, bottom panel) based on four subsets of TESS data. The error bars display the one-sigma error derived using \textit{SMURFS}.}
\label{fig:TESSAmpvar}
\end{center}
\end{figure}

As a final check, we used the TESS data set which is considerably shorter in time base, but has significantly less noise compared to the BRITE observations. Consequently, we split the light curve in the four blocks that are introduced by the visible gaps (see the top panel of Fig.~\ref{fig:TESS-lcs}). The blocks contain 8.24\,d, 10.0\d, 9.9\,d, and 10.5\,d of uninterrupted data with nearly 100\% of duty cycle. For each of the blocks, we determined the corresponding amplitudes of F$_1$ and F$_2$. The third independent pulsation frequency, F$_4$, has such a low amplitude that it does not appear in the frequency spectra of separate subsets. Hence, we had to discard it from our study of the pulsation amplitude behaviour. Figure~\ref{fig:TESSAmpvar} again illustrates that the amplitudes of F$_1$ and F$_2$ do not vary much during the 49.686 days of TESS observations even within one-sigma errors.

We conclude that no statistically significant variability of the pulsation amplitudes of F$_1$, F$_2$, and F$_4$ can be detected.

\subsection{Non-binarity of $\beta$\,Cas}
In the past, it was speculated several times that $\beta$\,Cas is actually a binary system. Although a detailed analysis by \citet{abt1965} illustrated the non-detection of a secondary component around $\beta$\,Cas, a recent study by \citet{liakos2017} again lists the star as a possible binary. Therefore, we applied the time delay method \citep{murphy2015} to our nearly four-years long photometric time series to check independently if a secondary can be detected.

The orbital movement of two components in a binary system introduces changes in the pulsation signal of $\delta$ Scuti pulsators throughout the orbit \citep{shibahashi2012}. This results in frequency modulation, where the amplitude spectrum shows additional peaks near the intrinsic pulsation frequencies and phase modulation, where the intrinsic pulsation signal arrives earlier (or later) depending on the orbital phase. This phase modulation is equivalent to the concept of light arrival time delays, i.e., the pulsation signal arrives later if the star is further away and vice versa. These time delays depend on the orbital parameters and it is therefore possible to constrain the orbit if time delays can be measured from the light curve \citep{murphy2015}.

We used \texttt{maelstrom} \citep{hey2020} to investigate our data sets for any time delay signal. Neither data set shows significant time delays. 
We therefore exclude a binary signal with $a \sin(i) / c \gtrsim 5$~s (see Fig. 8 of \citet{hey2020}) and periods $< 1000$~days.

\subsection{Evolutionary stage}
\begin{figure}
\begin{center}
\includegraphics[width=0.48\textwidth]{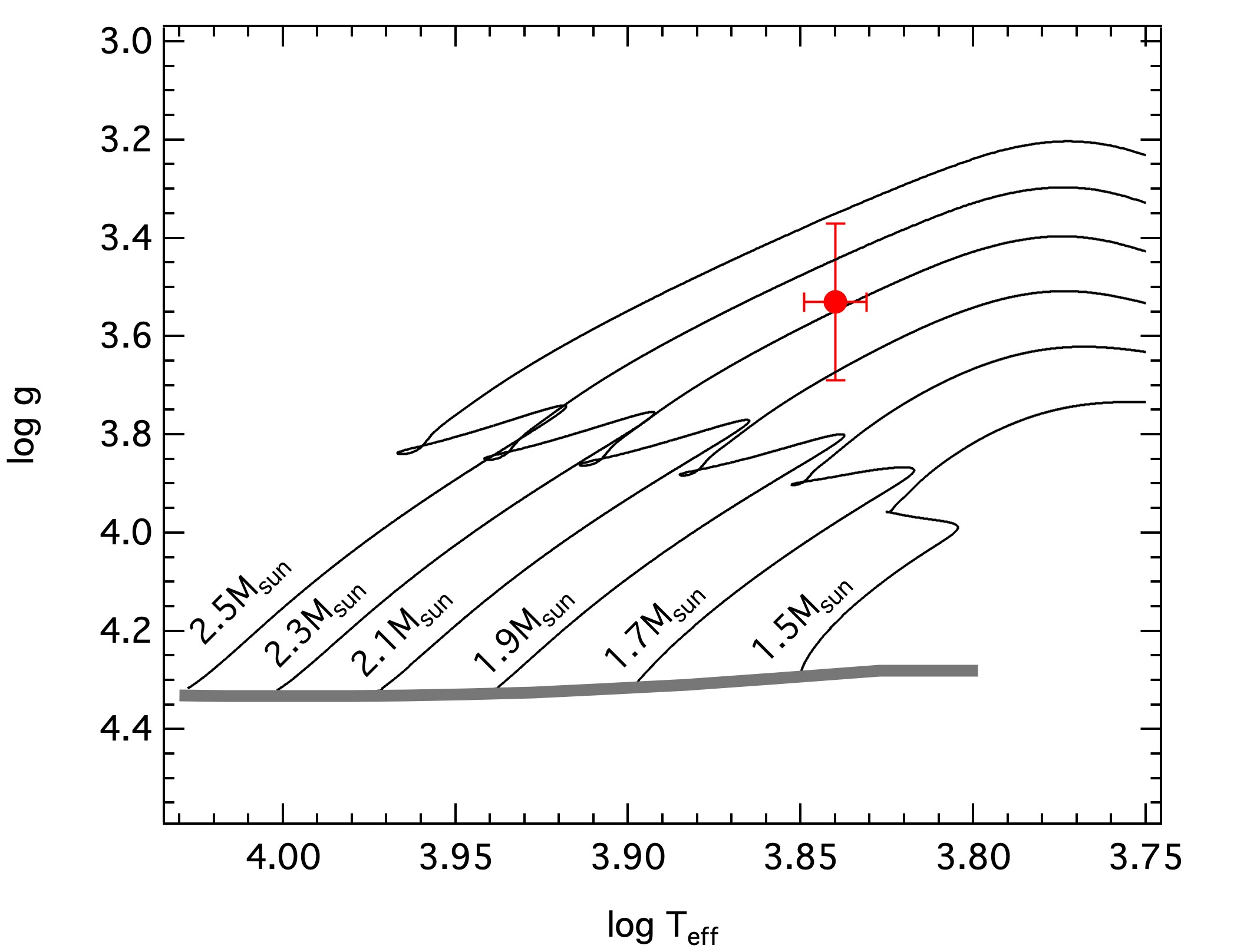}
\caption[]{Position of $\beta$\,Cas in the Kiel diagram based on our final adopted values for \Teff\ and \logg\ (red square). Post-main sequence tracks have been calculated using MESA version r-12115 \citep{paxton2011, paxton2013, paxton2015, paxton2018, paxton2019}.  The location of the zero-age main sequence (ZAMS) is given as a thick grey line.}
\label{fig:Kiel}
\end{center}
\end{figure}

Using the \Teff\ and \logg\ values determined from our analysis and the errors taken from the second approach described in Sect.~\ref{Sec:spectro} (i.e., the err$_2$ values in Table \ref{param}), the position of $\beta$\,Cas in the Kiel diagram is shown in Fig.~\ref{fig:Kiel}. From our analysis, we can confirm that $\beta$\,Cas' mass is around 2.1\,\Msun, and that it has to be a rather evolved star moving away from the TAMS. 

\subsection{Structure of the magnetic field}

Following the detection of a complex magnetic field on the surface of $\beta$\,Cas, we attempted to reconstruct the stellar field topology with the ZDI technique. The subset of 15 Stokes $V$ averaged LSD profiles suitable for this analysis extends over about two weeks, corresponding to approximately 14 stellar rotations. These data alone are insufficient for an unambiguous determination of the rotational period and are not suitable for a refined investigation of, for example, latitudinal differential rotation, temporal field evolution, or variation of the field structure with pulsational phase. Nevertheless, we found a plausible rotational period, compatible with the period range suggested by the previous interferometric analysis, which phases the Stokes $V$ profiles convincingly. Using this period we have successfully reproduced the circular polarisation observations with a fairly complex, yet static, magnetic field topology. The longitudinal field values follow a non-sinusoidal behavior, unlike dipolar fossil fields. Further high-cadence spectropolarimetric observations are required for probing the temporal evolution of this field and exploring its response to the global $\delta$~Scuti pulsations.

Finding such a complex surface magnetic field structure in a star with an effective temperature as high as 6920~K is very unusual. It is known that fossil magnetic fields exist in stars as cool as $T_{\rm eff}$\,$\approx$\,6500~K \citep{2003A&A...404..669K,2010A&A...520A..88S}. On the other hand the hottest stars with dynamo magnetic fields were previously found with a temperature up to $T_{\rm eff}$\,$\approx$\,6700~K  \citep{2014MNRAS.444.3517M,seach2020}. Judging by its complexity, the field of $\beta$\,Cas is almost certainly of dynamo origin. It is significantly more complex than any fossil fields of magnetic A, B, and O stars. Instead, its structure resembles the surface magnetic field topologies of well-studied cool, young, active stars such as AB~Dor \citep{1999MNRAS.302..437D}, V410~Tau \citep{2015A&A...580A..39K}, and Tap~26 \citep{2017MNRAS.467.1342Y}. Thus, our study establishes that the dynamo fields can exist in stars with $T_{\rm eff}$ of up to $\approx$\,6900~K. It is possible that the rapid rotation of $\beta$\,Cas, by increasing the thickness of the surface convection layer at the equator, facilitates the dynamo action, allowing it to exist to a higher $T_{\rm eff}$ than in slowly rotating stars.

%In any case, this object provides an interesting benchmark for theoretical modelling of dynamo processes in thin convective envelopes of F-type stars and for the study of the transition region between fossil and dynamo fields. 

Finally, about $\sim$10\% of O, B and A stars host a fossil field on the main sequence with a typical dipolar field strength of 3 kG. The strength of these fields at the stellar surface has been shown to decrease with time due to magnetic flux conservation as the stellar radius increases \citep{kochukhov2006,landstreet2007,shultz2019}, and probably to additional decrease due to, e.g., Ohmic decay. Since $\beta$\,Cas is evolved, we cannot exclude that it hosts such a fossil field that has reached a faint strength at the surface and is interacting with the dynamo field in the surface layer. Therefore, $\beta$\,Cas is possibly also an interesting target for the study of the evolution of fossil field and for the interaction between fossil and dynamo fields.

\section{Conclusions}

For several reasons, $\beta$\,Cas is an unusual star that combines several physical properties: \\
(i) It is a $\delta$ Scuti pulsator that shows only three independent {\it p}-mode pulsation frequencies even in multiple seasons of space photometry down to the few ppm-level. We suggest that its highest amplitude mode, F$_1$, can be identified as an $n=3$, $\ell = 2$, $m=0$ mode based on the pulsation constant and the ZDI analysis. No {\it g}-mode frequencies were detected.\\
(ii) It is one of the handful of $\delta$ Scuti stars known to date to show a measurable magnetic field. \\
(iii) Additionally, $\beta$\,Cas' magnetic field structure is quite complex, which is unusual in a star with an effective temperature as high as 6920~K.  According to its complexity, the field of $\beta$\,Cas is almost certainly of dynamo origin, which makes it the first $\delta$ Scuti object with a dynamo magnetic field. $\beta$\,Cas' rapid rotation may lead to a thicker surface convective layer and explain how a dynamo can exist.

Therefore, $\beta$\,Cas provides an interesting benchmark for the theoretical modelling of dynamo processes in thin convective envelopes of F-type stars and for the study of the transition region between fossil and dynamo fields.

\begin{acknowledgements}
KZ acknowledges support by the Austrian Fonds zur F\"orderung der wissenschaftlichen Forschung (FWF, project V431-NBL) and the Austrian Space Application Programme (ASAP) of the Austrian Research Promotion Agency (FFG). 
CN acknowledges support from PNPS (Programme National de Physique Stellaire). OK acknowledges research funding from the Swedish Research Council, the Swedish National Space Board, and the Knut and Alice Wallenberg foundation. This research has made use of the SIMBAD database operated at CDS, Strasbourg (France), of NASA's Astrophysics Data System (ADS), and of the VALD database, operated at Uppsala University, the Institute of Astronomy RAS in Moscow, and the University of Vienna. The research of A.F.J.M. has been supported by the Natural Sciences and Engineering Research Council (NSERC) of Canada.
Adam Popowicz was responsible for image processing and automation of photometric routines for the data registered by BRITE-nanosatellite constellation, and was supported by Silesian University of Technology Rector Grant 02/140/RGJ20/0001.
GAW acknowledges Discovery Grant Support from the National Science and Engineering Research Council (NSERC) of Canada.
\end{acknowledgements}

\bibliographystyle{aa} % style aa.bst
\bibliography{betcas} % your references Yourfile.bib

%%%%%%%%%%%%%%%%%%%%%%%%%%%%%%%%%%%%%%%%%%%%%%%%%%%%%%%%%%%
%%% APPENDIX
%%%%%%%%%%%%%%%%%%%%%%%%%%%%%%%%%%%%%%%%%%%%%%%%%%%%%%%%%%%
\begin{appendix}

\section{Additional BRITE and SMEI light curves}

\begin{figure*}
\begin{center}
\includegraphics[width=0.9\textwidth]{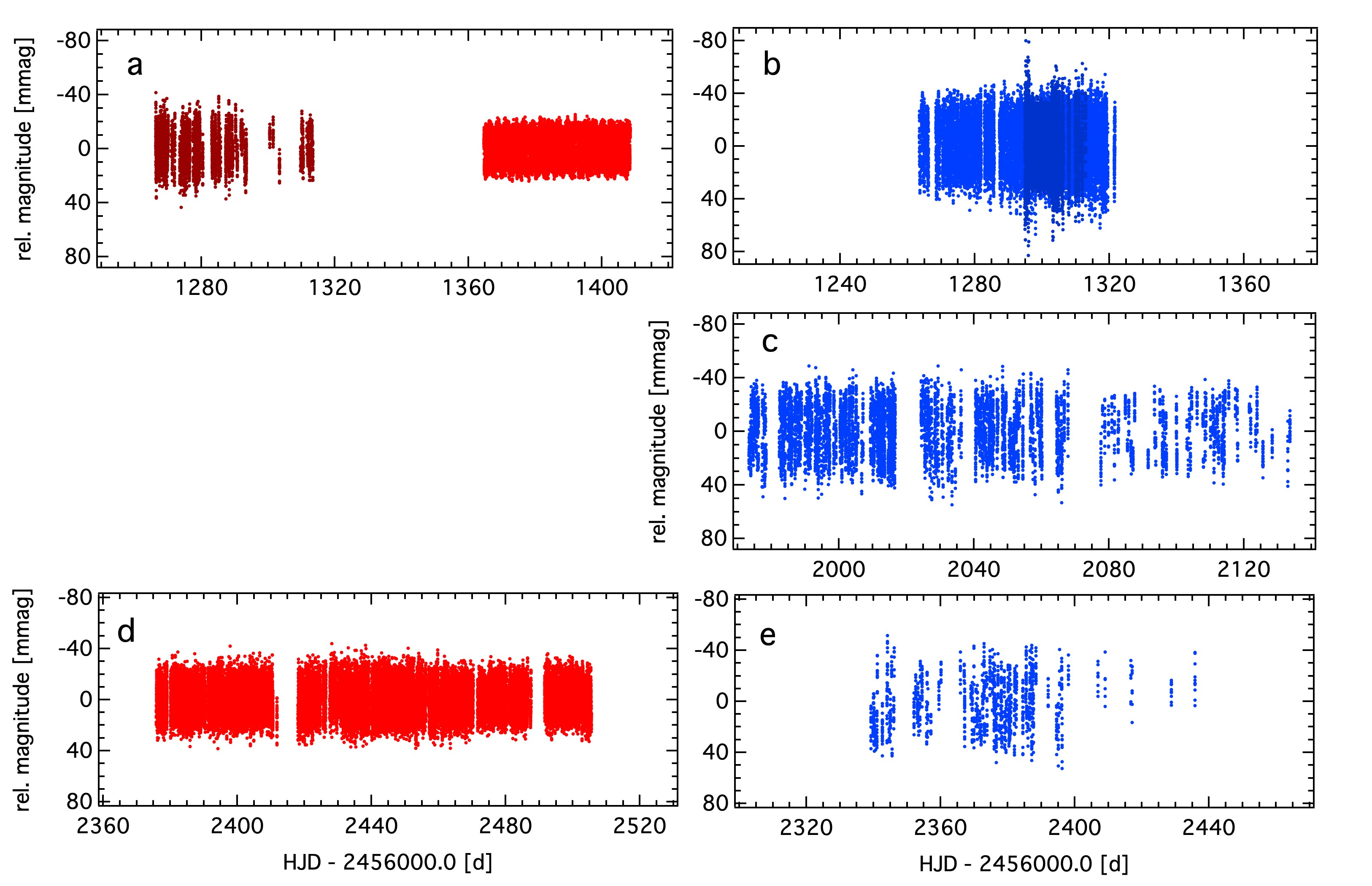}
\caption[]{BRITE photometric time series obtained by BTr (brighter red in panel a), BHr (darker red in panel a), BAb (brighter blue in panel b) and BLb (darker blue in panel b) in 2015, by BAb in 2017 (panel c), by BTr (panel d) and BAb (panel e) in 2018.}
\label{fig:BRITE-lcs-appendix}
\end{center}
\end{figure*}

The data obtained by BAb and BLb as well as the data obtained by BHr and BTr in 2015 were combined to one blue filter and one red filter light curve, respectively. Their properties are described in Table \ref{tab:obs}. As can be seen in panels a and b of Fig.~\ref{fig:BRITE-lcs-appendix}, observations of BAb and BLb were conducted simultaneously, while BHr and BTr observed $\beta$\,Cas at distinct times. Figure~\ref{fig:BRITE-lcs-appendix} also shows the BAb observations in 2017 (panel c) and the BTr and BAb observations in 2018 (panels d and e).

\begin{figure}
\begin{center}
\includegraphics[width=0.45\textwidth]{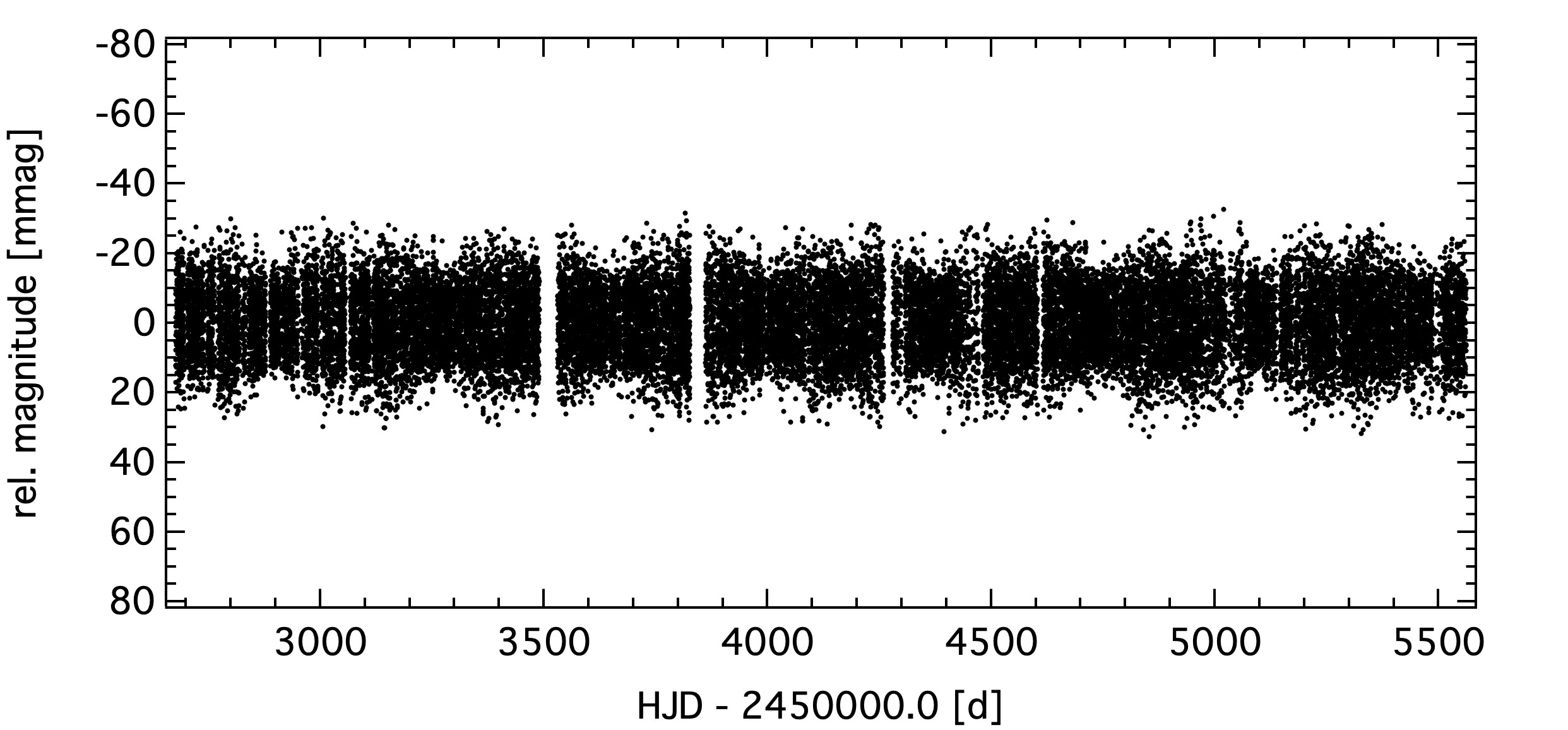}
\caption[]{SMEI photometric time series.}
\label{fig:SMEI-lc}
\end{center}
\end{figure}

Figure~\ref{fig:SMEI-lc} shows the full light curve obtained by SMEI.

\section{Amplitude spectra and spectral windows}
\label{app}

Below we provide additional figures illustrating the frequency analyses of the TESS and BRITE-Constellation 2015, 2017, and 2018 data as well as the spectral window functions for all data sets used in our analysis.

\begin{figure}
\begin{center}
\includegraphics[width=0.49\textwidth]{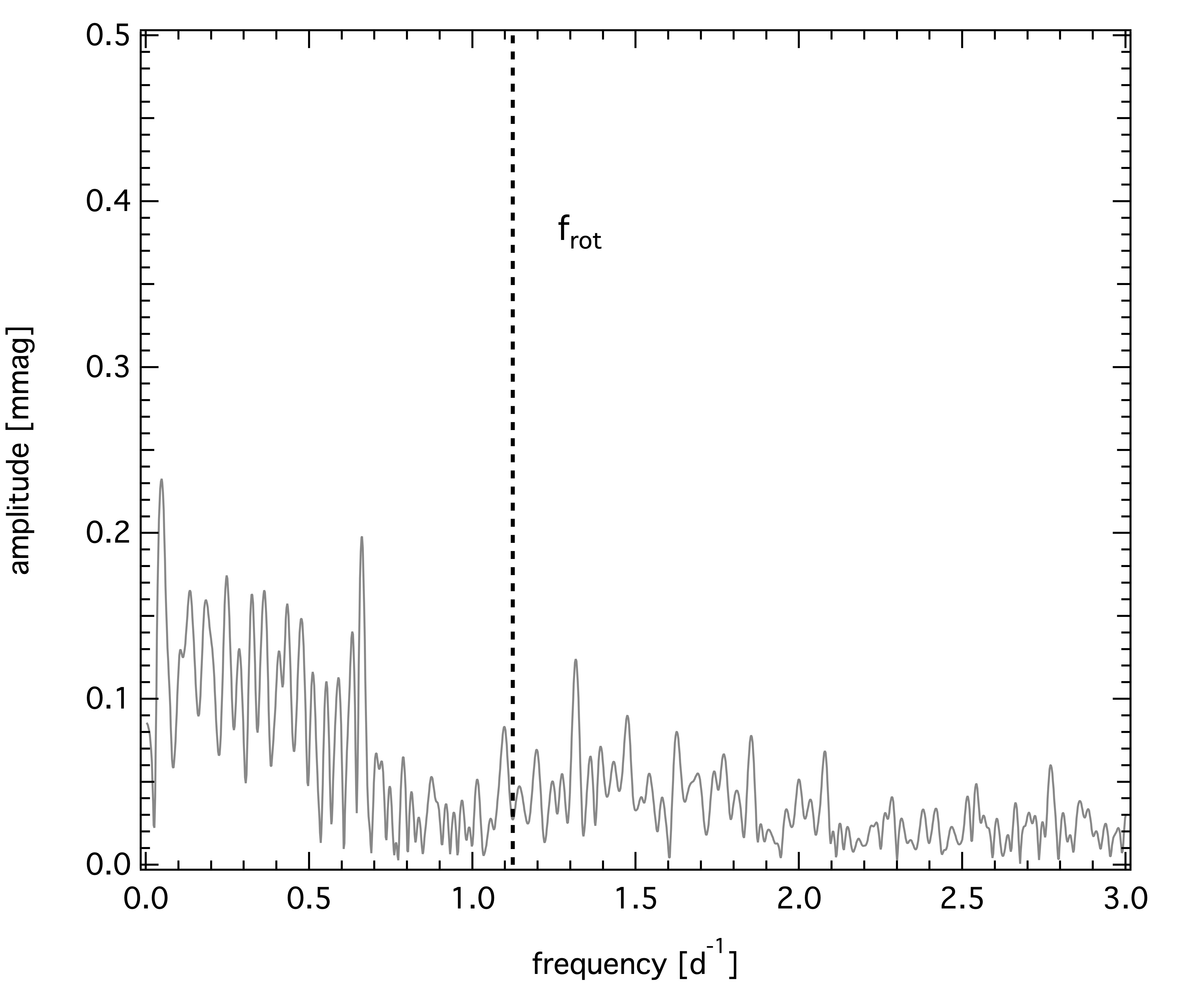}
\caption[]{Zoom into the low frequency domain of the amplitude spectrum of the original TESS data before application of a Gaussian filter. The position of the rotation frequency as determined by \citet{che2011} is indicated with a dashed line. }
\label{fig:TESS-zoom}
\end{center}
\end{figure}

Figure \ref{fig:TESS-zoom} shows a zoom into the low frequency domain of the amplitude spectrum of the original TESS data: the rotation frequency at 1.12\,\cd\ cannot be identified, there is increased noise in the region between 0.0 and 0.5\,\cd\, and no evidence for significant g-mode frequencies.

\begin{figure*}
\begin{center}
\includegraphics[width=0.9\textwidth]{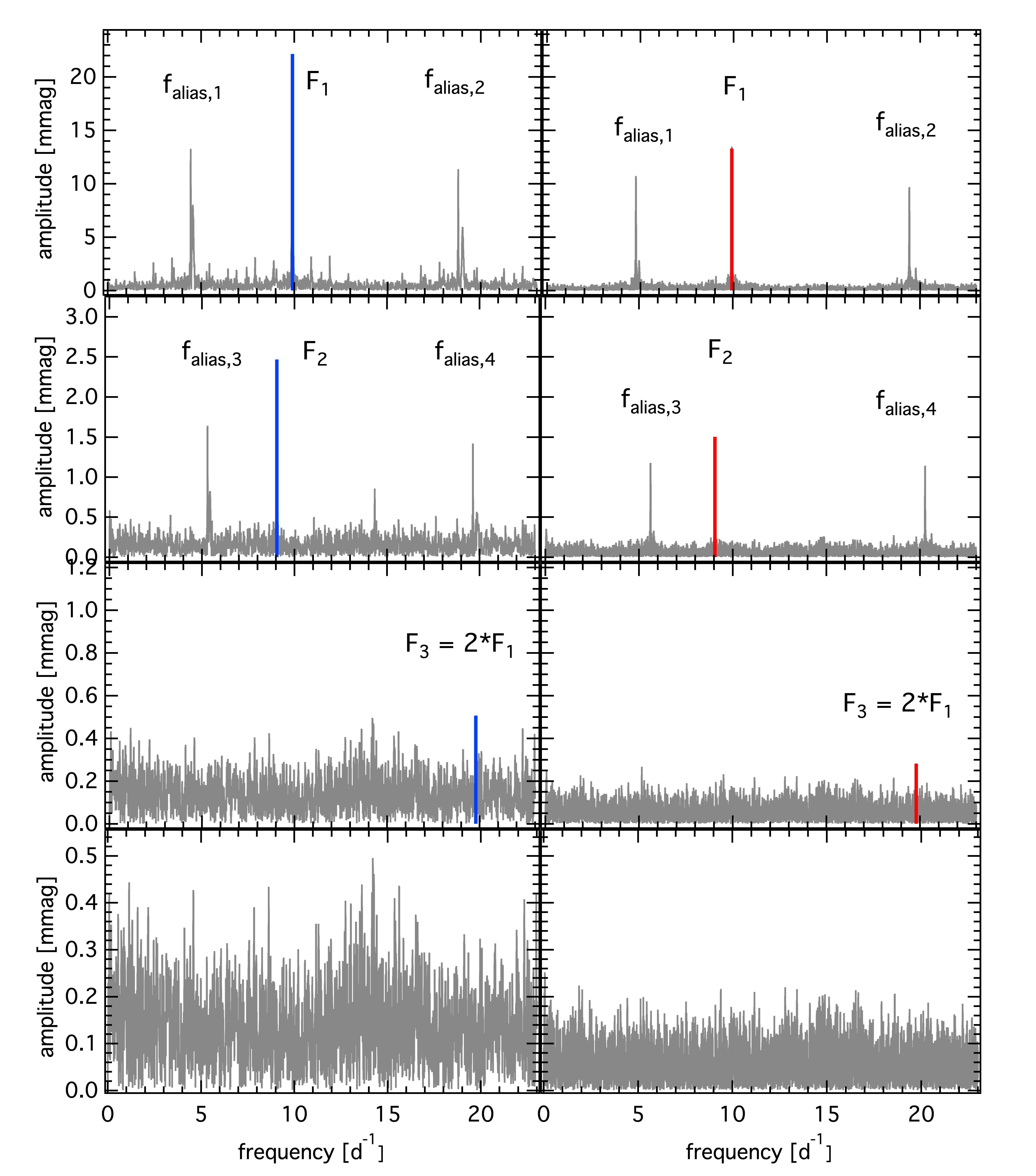}
\caption[]{Amplitude spectra of the BRITE-Constellation data obtained in 2015: Combined blue filter data are shown on the left side, combined red filter data on the right side. 
Top panels show the amplitude spectra of the original data with F$_1$ identified, the second panels illustrate the amplitude spectra after prewhitening F$_1$, the third panels those after prewhitening F$_1$ and F$_2$, and the bottom panel displays the residual amplitude spectra after prewhitening all significant frequencies. An explanation for the identified alias frequencies is given in the text.}
\label{amps2015}
\end{center}
\end{figure*}

%\begin{figure}
%\begin{center}
%\includegraphics[width=0.5\textwidth]{betaCas2015_residuals.jpg}
%\caption[]{Residual amplitude spectra for the 2015 combined blue data (in blue pointing upwards) and combined red data (in red pointing downwards) after prewhitening all significant frequencies.}
%\label{residuals2015}
%\end{center}
%\end{figure}

\begin{figure}
\begin{center}
\includegraphics[width=0.5\textwidth]{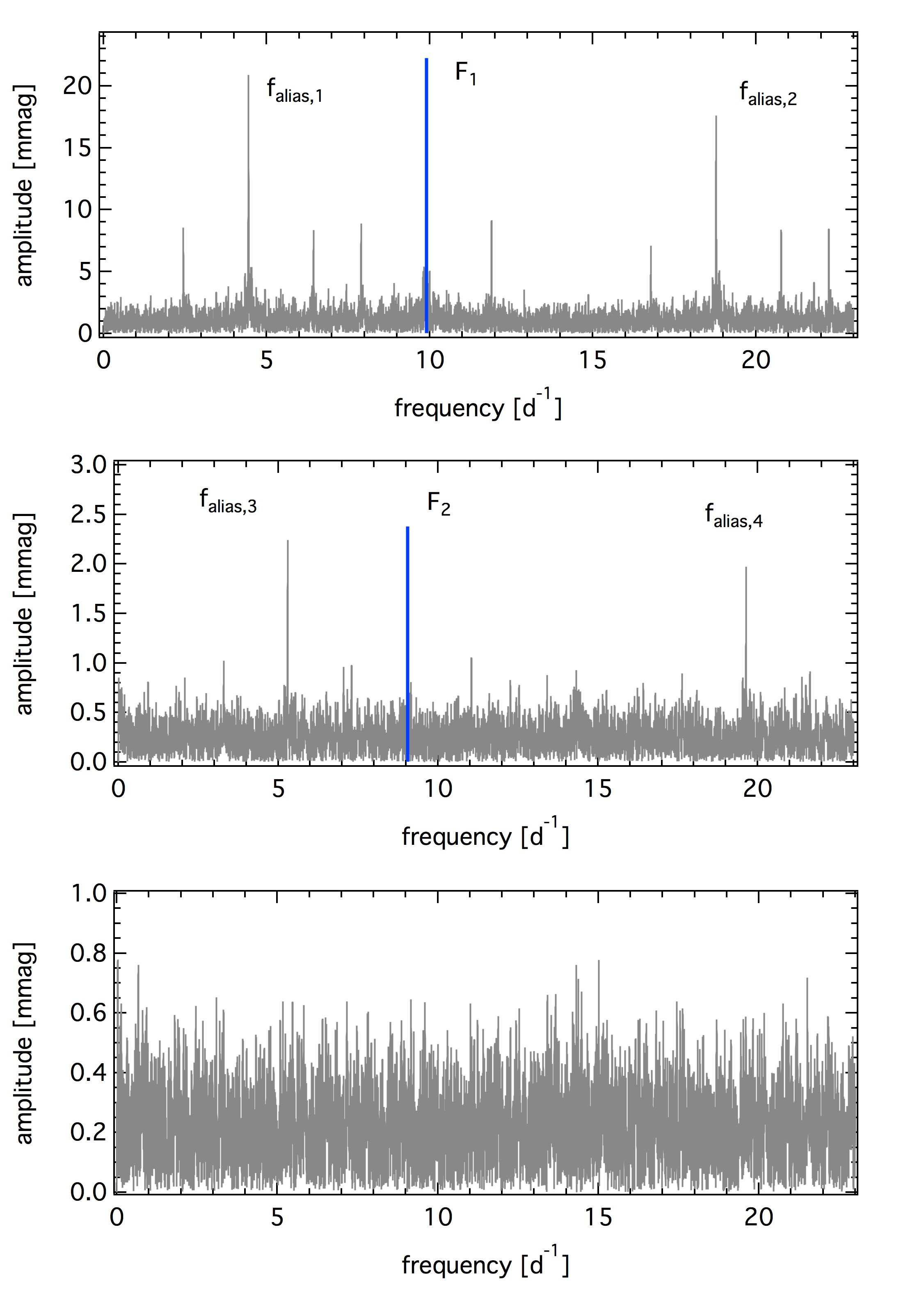}
\caption[]{Amplitude spectra of the BRITE-Constellation data obtained in 2017 with BAb only: the top panel shows the amplitude spectrum of the original data with F$_1$ identified, the middle panel illustrates the amplitude spectrum after prewhitening F$_1$, and the bottom panel shows the residuals after prewhitening with the significant frequencies. Note that the noise of the 2017 BAb data is significantly higher than for all the other combined data sets, hence F3 remains hidden in the noise.}
\label{amps2017}
\end{center}
\end{figure}

\begin{figure*}
\begin{center}
\includegraphics[width=0.9\textwidth]{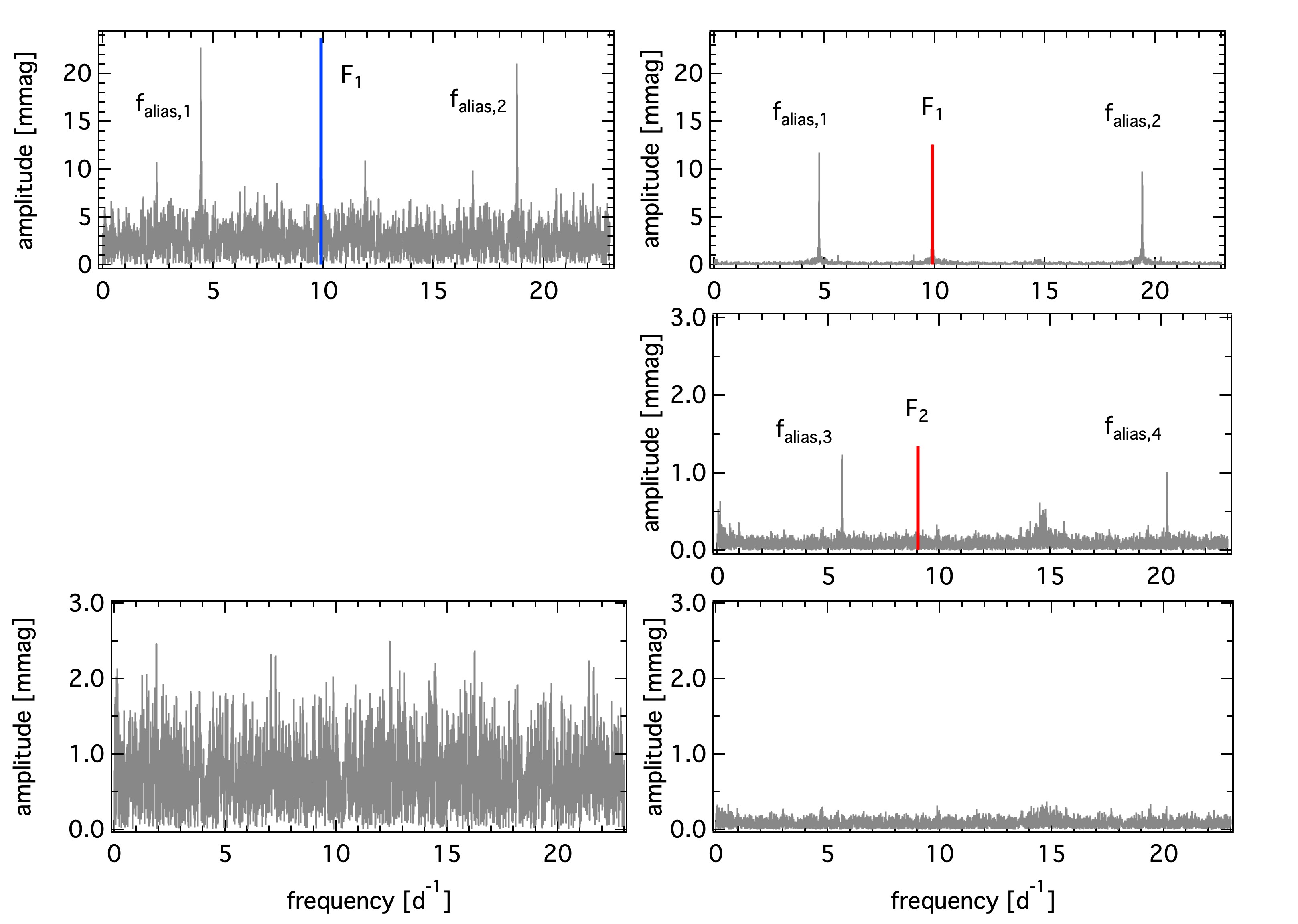}
\caption[]{Amplitude spectra of the BRITE-Constellation data obtained in 2018: BAb data are shown on the left side, BTr data on the right side. 
Top panels show the amplitude spectra of the original data with F$_1$ identified, the middle panel (BTr only) illustrates the amplitude spectra after prewhitening F$_1$, and the bottom panels the residuals after prewhitening all respective significant frequencies. An explanation for the identified alias frequencies is given in the text.}
\label{amps2018}
\end{center}
\end{figure*}

\begin{figure*}[htb]
\begin{center}
\includegraphics[width=0.9\textwidth]{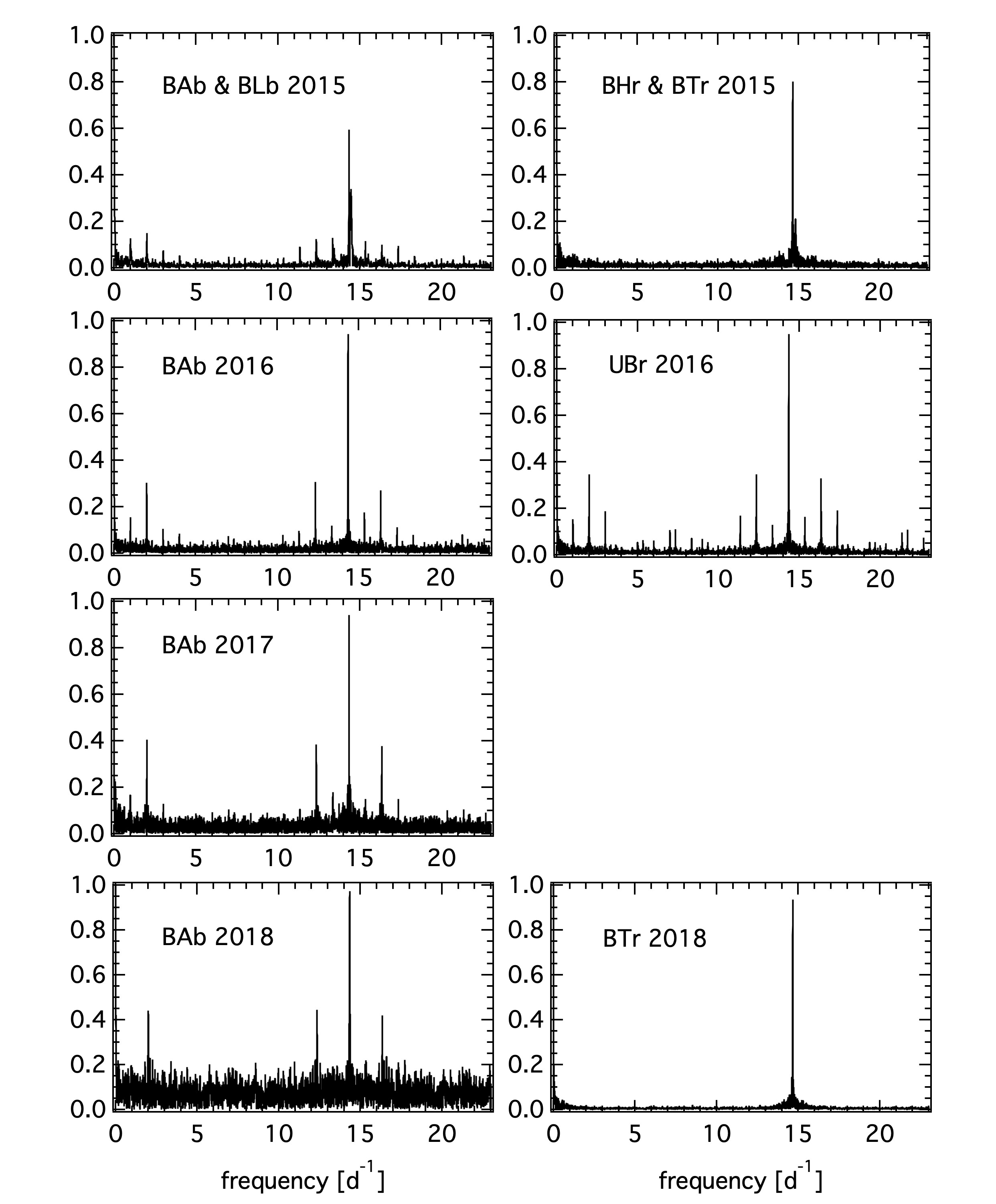}
\caption[]{Spectral window functions for the BRITE-Constellation data obtained in 2015, 2016, 2017, and 2018 in the blue (left panels) and the red filter (right panels).}
\label{BRITE-spws}
\end{center}
\end{figure*}

\begin{figure}
\begin{center}
\includegraphics[width=0.5\textwidth]{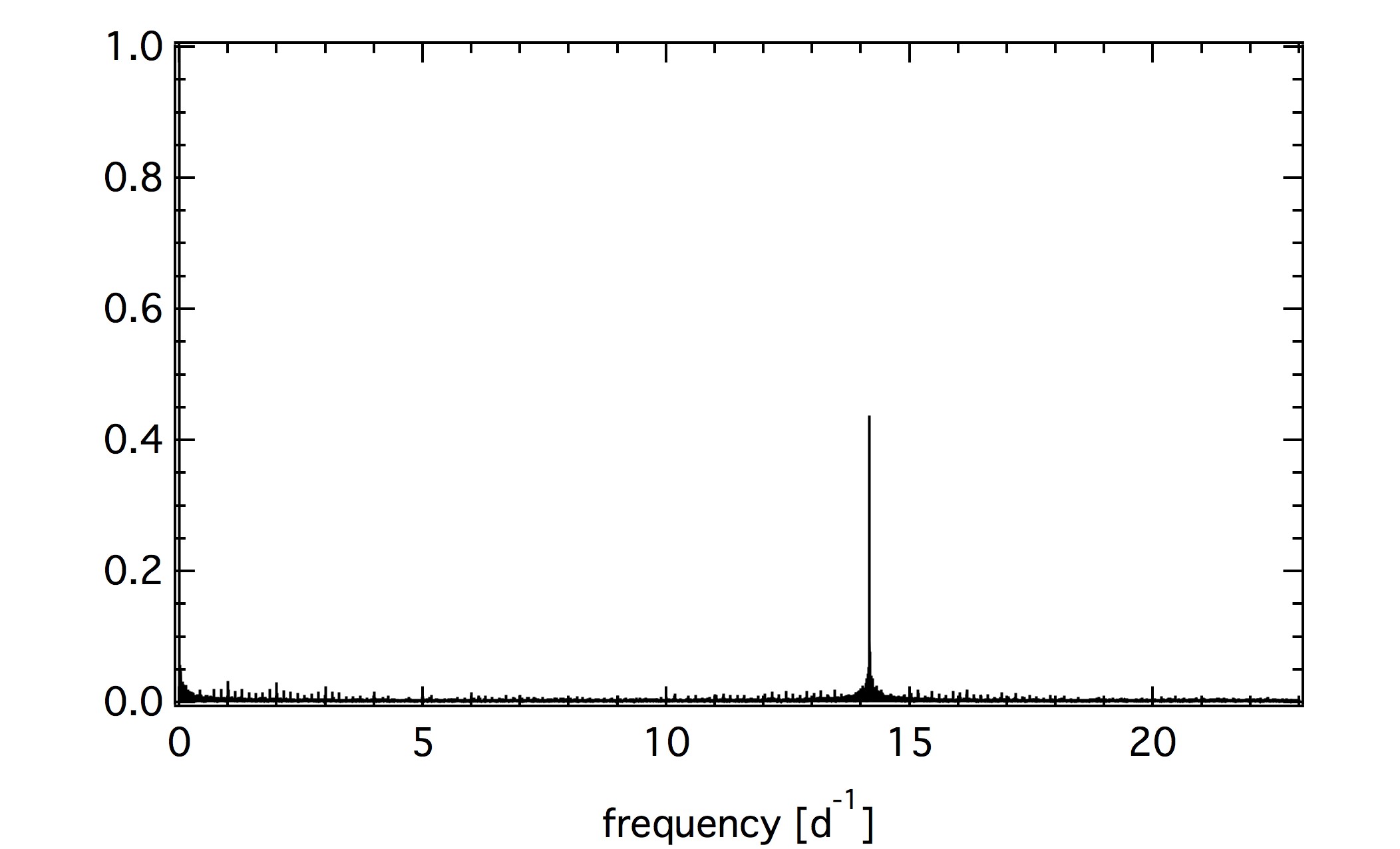}
\caption[]{Spectral window function for the SMEI data.}
\label{SMEI-spw}
\end{center}
\end{figure}

\begin{figure}
\begin{center}
\includegraphics[width=0.5\textwidth]{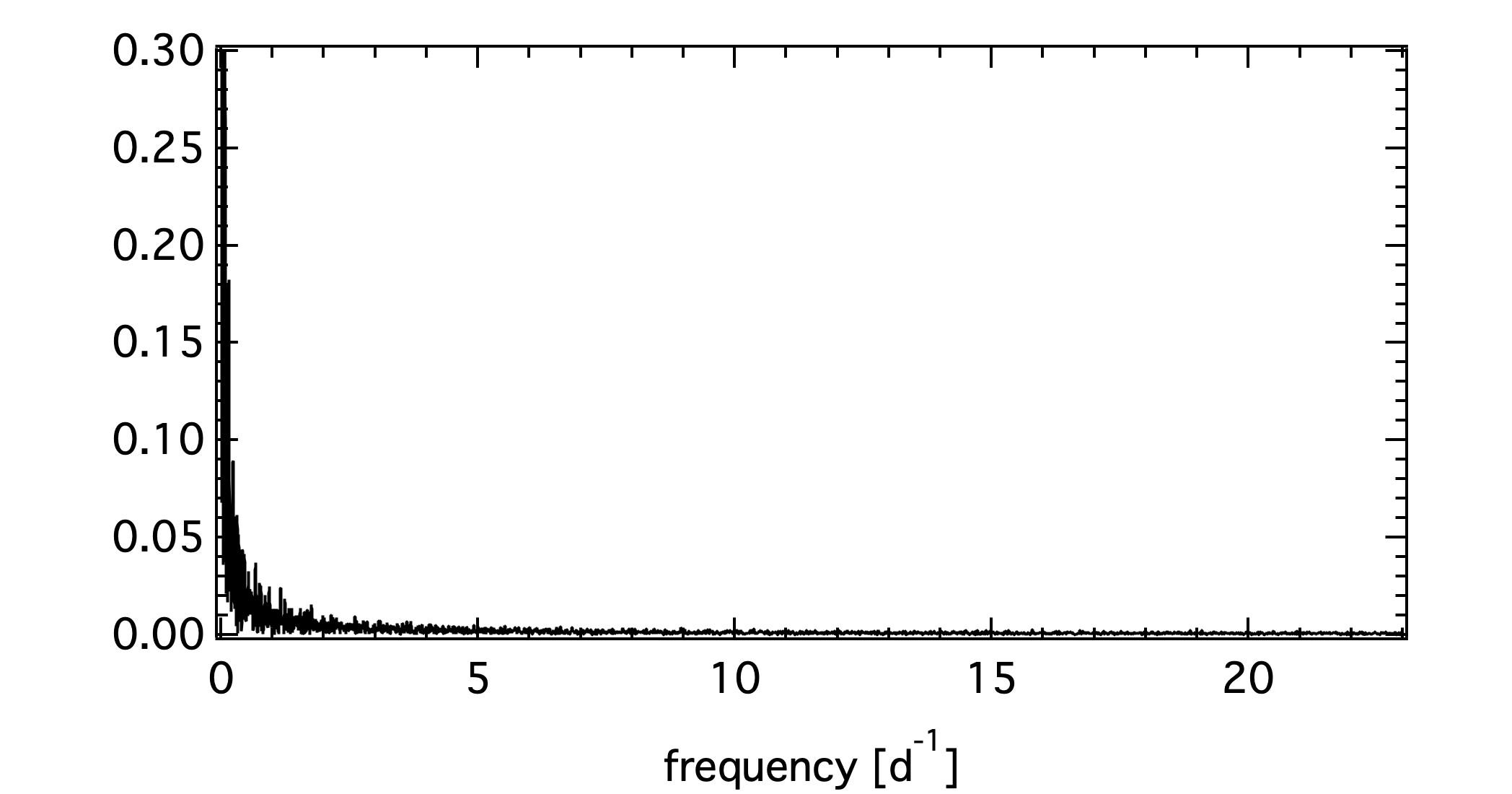}
\caption[]{Spectral window function for the TESS data. Note the different Y-axis scale compared to the spectral window functions of the BRITE and SMEI data.}
\label{TESS-spw}
\end{center}
\end{figure}

\section{LSD profiles}
\label{all_lsd}

Here we illustrate all LSD Stokes I, V, and null profiles of $\beta$\,Cas derived from the Narval observations obtained in 2013, 2014, and 2015.
\begin{figure*}
\begin{center}
\includegraphics[width=0.8\textwidth]{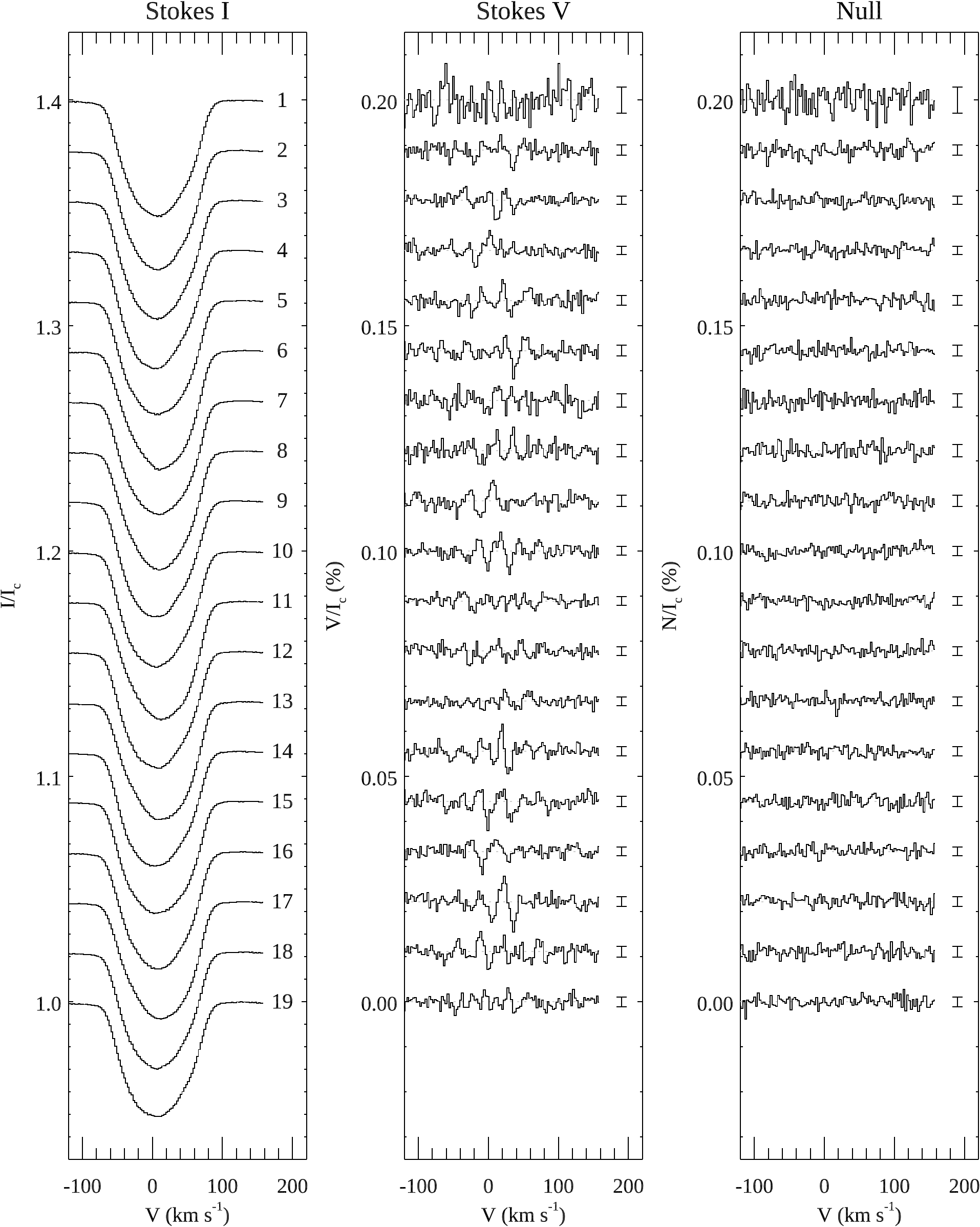}
\caption[]{Observed LSD Stokes I (left), V (middle), and null (right) profiles of $\beta$\,Cas. Spectra are shifted vertically according to the observing date, from the earliest on top to the latest at the bottom. The numbering of profiles in the Stokes I panel corresponds to the polarimetric measurement number given in Table~\ref{logNarval}.}
\label{all-lsd}
\end{center}
\end{figure*}

\section{Distribution of magnetic energy over spherical harmonic modes}

\begin{figure}
\begin{center}
\includegraphics[width=\hsize]{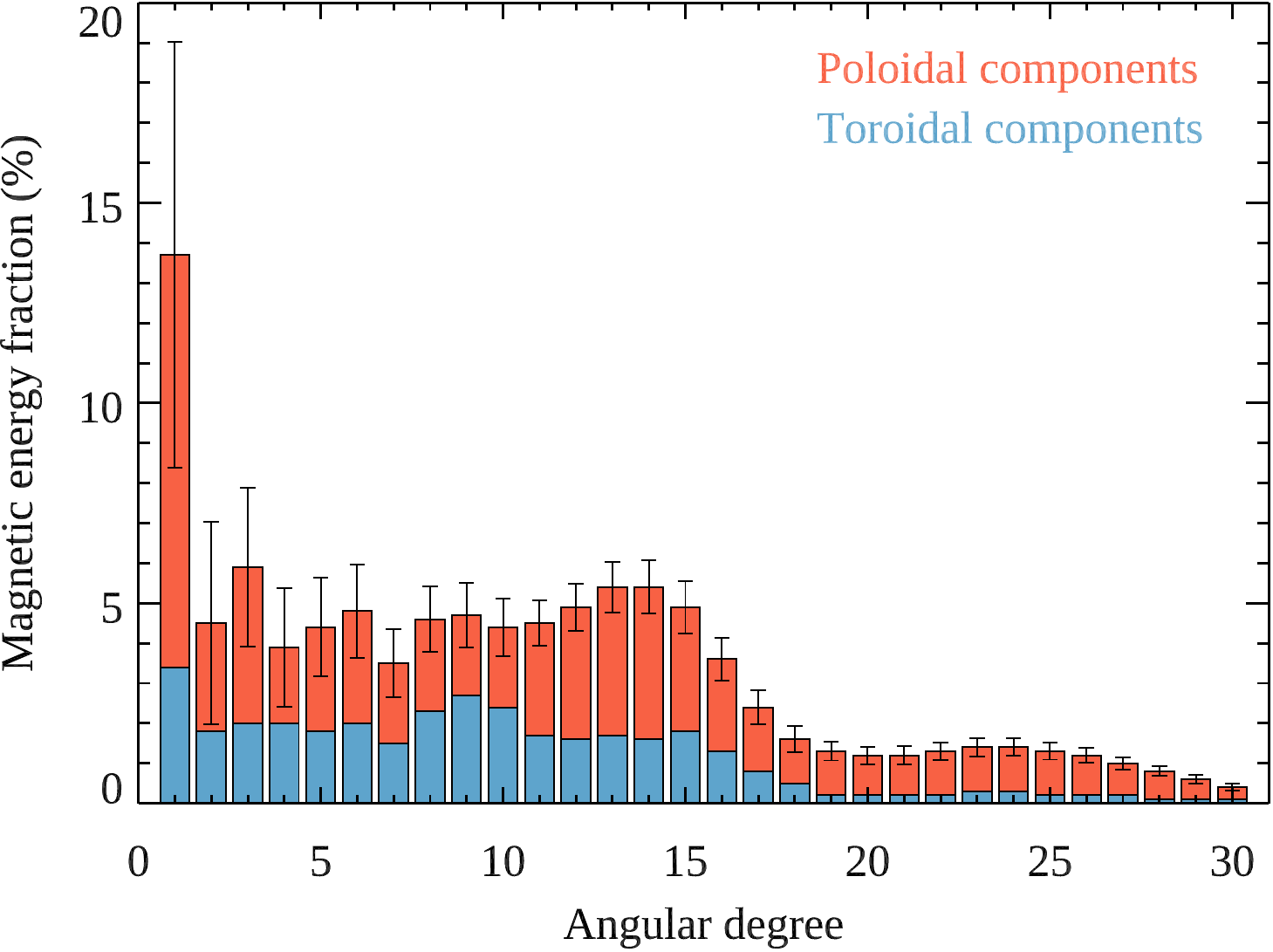}
\caption[]{Magnetic field energy as a function of angular degree $\ell$ of the spherical harmonic expansion for the magnetic field topology recovered with ZDI. Different colours show contributions of the poloidal and toroidal components. Error bars correspond to the uncertainties estimated as described in Sect.~\ref{sect:ZDI}.}
\label{zdi-ell}
\end{center}
\end{figure}

\end{appendix}

\end{document}